\documentclass[aps,nofootinbib,showpacs,preprintnumbers,amsmath,amssymb]{revtex4}
\def\be{\begin{equation}}
\def\ee{\end{equation}}
\def\bea{\begin{eqnarray}}
\def\eea{\end{eqnarray}}
\newcommand{\A}{{\mathcal{A}}}
\newcommand{\tA}{{\widetilde {\mathcal{A}}}}
\newcommand{\ta}{{\widetilde a}}
\newcommand{\ntH}{{\mathfrak A}}
\newcommand{\tH}{{\widetilde {\mathfrak A}}}
\newcommand{\tk}{{\widetilde k}}
\newcommand{\tZ}{{\widetilde Z}}
\newcommand{\tB}{{\widetilde B}}

\newcommand{\bL}{\overline{\Lambda}}

\usepackage{graphics}
\usepackage{graphicx}
\usepackage{dcolumn}
\usepackage{bm}
\usepackage{epsfig}
\usepackage{graphicx}
\usepackage{multirow}

\begin{document}

\vspace{1cm}

\preprint{USM-TH-289, arXiv:1106.4275v3, published in: J.Phys.G, G39 (2012) 065005}

\title{Analogs of noninteger powers in general analytic QCD\footnote{
The changes in comparison with arXiv:1106.4275v2: section II is extended (after Eq.(11));
in section III: the part between eqs.(36) and (38) is new; appendix B is new; the part
between eqs.(44) and (50) is new; the part between eq.(54) and the end of 
section III is new.}}

\author{Gorazd Cveti\v{c}}
 \email{gorazd.cvetic@usm.cl}
\affiliation{Department of Physics and Centro Cient\'{\i}fico-Tecnol\'ogico de Valpara\'{\i}so,
Universidad T\'ecnica Federico Santa Mar\'{\i}a, Casilla 110-V, 
Valpara\'{\i}so, Chile}

\author{Anatoly V.~Kotikov} 
 \email{kotikov@theor.jinr.ru}
\affiliation{Bogolyubov Laboratory of Theoretical Physics, Joint Institute for 
Nuclear Research, 141980 Dubna, Russia}  
 
\begin{abstract}
In contrast to the coupling parameter in the usual perturbative QCD (pQCD), 
the coupling parameter in the analytic QCD models has cuts only on the 
negative semiaxis of the $Q^2$-plane (where $q^2 \equiv - Q^2$ is the momentum 
squared), thus reflecting correctly the analytic structure of the spacelike 
observables. 
The Minimal Analytic model (MA, named also APT) 
of Shirkov and Solovtsov removes the 
nonphysical cut (at positive $Q^2$) of the usual pQCD coupling and keeps the 
pQCD cut discontinuity of the coupling at negative $Q^2$ unchanged. In order 
to evaluate in MA the physical QCD quantities whose perturbation expansion 
involves noninteger powers of the pQCD coupling, a specific method of 
construction of MA analogs of noninteger pQCD powers was developed by Bakulev, 
Mikhailov and Stefanis (BMS). We present a construction, applicable now 
in any analytic QCD model, of analytic analogs of noninteger pQCD powers; 
this method generalizes the BMS approach obtained in the framework of MA. 
We need to know only the discontinuity function of the analytic coupling 
(the analog of the pQCD coupling) along its cut in order to obtain the analytic
analogs of the noninteger powers of the pQCD coupling, as well as their timelike 
(Minkowskian) counterparts. As an illustration, we apply the method to the 
evaluation of the width for the Higgs decay into $b {\bar b}$ pair.
\end{abstract}

\maketitle

\section{Introduction}
\label{sec:intr}  

It is well known that the perturbative approach to QCD (pQCD),
while working well in evaluation of physical quantities at high
momentum transfer ($|q^2| \agt 10^1 \ {\rm GeV}^2$), becomes increasingly
unreliable at low momenta ($|q^2| \sim 1 \ {\rm GeV}^2$). One of the main
reasons for this is the singularity structure of the pQCD
coupling parameter $a_{\rm pt}(Q^2) \equiv \alpha_s(Q^2)/\pi$ 
at spacelike low momenta $q$:
$(0 <)$ $Q^2 \equiv - q^2  \sim 1 \ {\rm GeV}^2$. This singularity structure
does not reflect correctly the analyticity structure of the
(to be evaluated) spacelike observables ${\cal F}(Q^2)$. The latter,
by the general principles of the (local) 
quantum field theory \cite{BS,Oehme},
must be analytic functions in the entire $Q^2$ plane except
on the cut on the negative semiaxis: 
$Q^2 \in \mathbb{C} \backslash (-\infty, 0]$.
Qualitatively the same analytic properties should have 
also the coupling parameter $\A_1(Q^2)$ that is used (instead of
$a_{\rm pt}(Q^2)$) to evaluate the spacelike observables ${\cal F}(Q^2)$.

\begin{figure}[htb]
\centering\epsfig{file=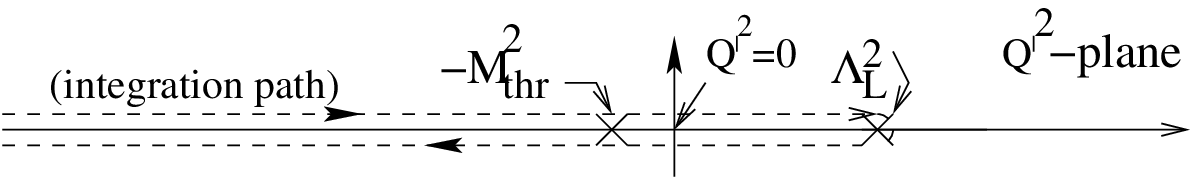,height=.06\textheight}
\vspace{-0.4cm}
\caption{The path of integration for the integral (\ref{anptdisp}).}
\label{integrpath}
\end{figure}

The first such analytic version was constructed in 
\cite{ShS,MSS,Sh}, where the discontinuity function
of pQCD $\rho_1^{\rm (pt)}(\sigma) = {\rm Im} a_{\rm pt}(Q^2=-\sigma - i \epsilon)$
was kept unchanged on the entire negative axis in the $Q^2$-plane.
More specifically, the use of the Cauchy theorem for the (powers
of the) pQCD coupling gives
\begin{equation}
a^n_{\rm pt}(Q^2) = 
\frac{1}{\pi} 
\int_{\sigma= - \Lambda_{\rm L}^2 - \eta}^{\infty}
\frac{d \sigma \ {\rm Im}  a^n_{\rm pt}(-\sigma-i \epsilon)}{(\sigma + Q^2)},
\label{anptdisp}
\end{equation}
where the integration is along the entire cut of the pQCD coupling
in the $Q^2$-plane $(-\infty,+\Lambda_{\rm L}^2)$, with $0< \Lambda_{\rm L}^2
\sim 1 \ {\rm GeV}^2$ being the (Landau) branching point, and $\eta \to +0$
(see fig.~\ref{integrpath}). 
Elimination of the unphysical (Landau) cut $(0,+\Lambda_{\rm L}^2)$
in the above dispersion relation leads to the aforementioned
Minimal Analytic (MA)\footnote{Another, somewhat different, approach, performs
minimal analytization of $d \ln a_{\rm pt}(Q^2)/ d \ln Q^2$ function,
\cite{Nesterenko}. An analytization using Borel transform of
observables and the minimal analytization approach, 
was presented in \cite{Karanikas}.
For reviews of various types of analytic 
QCD models, see \cite{Prosperi,Shirkov,Cvetic,Bakulev}.} coupling
\begin{equation}
{{\A_n^{\rm {(MA)}}}}(Q^2) = \frac{1}{\pi} \int_{\sigma= 0}^{\infty}
\frac{d \sigma \ {\rm Im}  
{a^n_{\rm pt}(-\sigma-i \epsilon)}}{(\sigma + Q^2)} \quad (n=1,2,\ldots).
\label{MAAndisp}
\end{equation}
It is named also Analytic Perturbation Theory (APT), \cite{ShS,MSS,Sh}.
It is applied usually in the ${\overline {\rm MS}}$ renormalization scheme, or in
truncated versions of that scheme. 
The method of equation (\ref{MAAndisp}) 
allows us to evaluate the MA coupling analogs
$\A_{n}^{\rm (MA)}(Q^2)$ of the pQCD coupling powers
$a_{\rm pt}^{n}(Q^2)$ even when $n$ is noninteger $(n \mapsto \nu$). 

MA gets its only free parameter, the QCD scale ${\overline \Lambda}$, 
fixed by the requirement that it reproduce high energy QCD quantities 
($|Q^2|  \stackrel{>}{\sim} 10^1 \ {\rm GeV}^2$), and in this regime it gives good results,
\cite{Sh}. It gives good results for the Bjorken polarized sum rule
(a spacelike quantity) even at low $Q^2$, \cite{BjPSR,Bjnew}, although
at very low $Q^2 \approx 0.1 \ {\rm GeV}^2$ it apparently requires a modification
\cite{Bjnew}.
Also due to duality violations we should expect that, 
at low $\sigma \sim 1 \ {\rm GeV}^2$, the discontinuity function 
$\rho_1(\sigma) \equiv {\rm Im} \A_1(-\sigma - i \epsilon)$
in analytic QCD models will deviate significantly from the pQCD counterpart
$\rho_1^{\rm (pt)}(\sigma) \equiv {\rm Im} a_{\rm pt}(-\sigma - i \epsilon)$.
Another reason for the need of such a deviation is the apparent inability
of MA to reproduce the correct value of the well-measured (timelike) low-energy QCD 
observable $r_\tau$, the strangeless semihadronic decay ratio of the $\tau$ lepton. 
Its present-day experimental value is $r_\tau({\rm exp.}) = 0.203 \pm 0.004$, 
 \cite{ALEPH,DDHMZ}.\footnote{$r_{\tau}$ represents the QCD part of the
decay ratio $R_{\tau}$, i.e., $(r_{\tau})_{\rm pt} = a_{\rm pt} + {\cal O}(a_{\rm pt}^2)$.}
In MA, the predicted values are in the range of $0.13$-$0.14$, \cite{MSS,MSSY},
unless the values of the current masses of the light quarks ($m_u$, $m_d$, $m_s$) are 
abandoned and effective quark masses $\approx 0.25$-$0.45$ GeV 
are used instead \cite{Milton:2001mq}.
This numerical loss in the size of $r_\tau$ in MA appears to be connected 
with the elimination of the unphysical (Euclidean) part of the branch cut 
contribution of perturbative QCD, while keeping the discontinuity along the
rest of the cut unchanged \cite{Geshkenbein:2001mn}. 
Furthermore, perturbative QCD models which are 
simultaneously also analytic (anpQCD), have also been investigated,
\cite{CKV}, and they turn out to give too low $r_{\tau}$ value
($r_{\tau} < 0.16$) unless their beta-function is modified in such a
manner as to give convergence only in the first four terms of expansion,
followed by explosive growth in the subsequent terms due to a rather singular
choice of the renormalization scheme.

Therefore, in general analytic QCD models, we must expect 
the following form of the dispersion relation: 
\be
\A_1(Q^2) 
= \frac{1}{\pi} \int_{M_{\rm thr}^2}^{+\infty} \ d \sigma 
\frac{ \rho_1(\sigma) }{(\sigma + Q^2)} \ ,
\label{dispA1}
\ee
with $\rho_1(\sigma) \equiv {\rm Im} \A_1(-\sigma - i \epsilon)$ deviating
from $\rho_1^{\rm (pt)}(\sigma)$ at low $\sigma$, and 
$(0 <)$ $M_{\rm thr}^2 \sim M_{\pi}^2$ being a threshold mass of
the cut $(-\infty, -M_{\rm thr}^2)$ of $\A_1(q^2)$ in the $Q^2$-plane.
Having in such general analytic QCD models no direct relation 
of $a_{\rm pt}(Q^2)$ with $\rho_1(\sigma)$, the general method
of obtaining $\A_n(Q^2)$, the analytic analog of the power $a_{\rm pt}(Q^2)^n$,
is not as straightforward as in equation (\ref{MAAndisp}) in the
case of MA. In \cite{CV1,CV2}, the higher power 
analogs $\A_n(Q^2)$ for integer $n$'s and for any analytic QCD
were constructed as linear combinations of logarithmic
derivatives $\tA_k(Q^2) \propto d^{k-1} \A_1(Q^2)/d (\ln Q^2)^{k-1}$
($k \geq n$),\footnote{The relations between $\tA_k$'s and $\A_n$'s
allowing for recurrent construction of $\A_n$, for integer $n$, 
were given also in \cite{Shirkov:2006nc,Shirkov},
within the context of the MA model of \cite{ShS,MSS,Sh}.}
such that the evaluation of observables
in analytic QCD leads to suppressed dependence of the
evaluated truncated analytic series when the number of
the terms in the series is increased. For the construction
of $\tA_n(Q^2)$ (and its timelike conterpart $\tH_n(\sigma)$),
only the knowledge of $\rho_1(\sigma)$ (or equivalently,
of $\A_1(Q^2)$) is needed. The construction of higher power 
analogs $\A_n$, not as powers of $\A_1$ 
but rather as linear operations on $\A_1$, 
has an attractive functional feature: it is compatible
with linear integral transformations (such as Fourier or Laplace) 
\cite{Shirkov:1999np}.

It turns out that some observables, in particular mass-dependent
ones, have pQCD expansion which involves noninteger
powers $a_{\rm pt}(Q^2)^{\nu}$. In order to evaluate such observables
in any analytic QCD, we need to construct their analytic
analogs $A_{\nu}(Q^2)$ (and their  timelike conterparts $\ntH_{\nu}(\sigma)$).
In the case of MA, a method of calculating such quantities
was developed and applied in \cite{BMS1,BMS2,BMS3,BP}
(for a review, see \cite{Bakulev}), 
their method being different from the direct evaluation
(\ref{MAAndisp}) with $n \mapsto \nu$ ($\nu$ noninteger). The analytic
properties of their couplings $\A_{\nu}^{\rm (MA)}(Q^2)$ can be
seen more clearly than in the formulas (\ref{MAAndisp}), but
numerically they are equivalent. In the present paper, we 
present the method of construction of  $\A_{\nu}(Q^2)$'s that is
applicable in any analytic QCD model.
Below we will demonstrate that,
within the MA model of Shirkov and Solovtsov,
our approach gives the same result
as the approach of \cite{BMS1,BMS2,BMS3} 
in the leading (one-loop) order of perturbation theory.
Above the leading order in MA, 
the results of \cite{BMS1,BMS2,BMS3} are represented
as certain expansions via the leading order results. Such types of expansions 
are absent in our approach.

In sections \ref{sec:lognonder} and \ref{sec:man} 
we derive the spacelike analogs
$\A_{\nu}(Q^2)$ of noninteger pQCD powers $a_{\rm pt}(Q^2)^{\nu}$,
as functions of $\rho_1(\sigma)$ of the analytic QCD model.
In sec.~\ref{sec:lognonder}, the construction leads us first to 
(noninteger counterparts)
of the logarithmic derivatives, $\ta_{{\rm pt}, \nu}(Q^2)$,
and their analytic analogs $\tA_{\nu}(Q^2)$.
In sec.~\ref{sec:man} we relate $\ta_{{\rm pt}, \nu}(Q^2)$
with the pQCD (noninteger) powers $a_{\rm pt}(Q^2)^{\nu+m}$;
this relation is derived in appendix \ref{App1}. This allows
us to evaluate the mentioned spacelike observables ${\cal F}(Q^2)$
in analytic QCD using either the analytic analogs $\tA_{\nu}(Q^2)$
of $\ta_{{\rm pt}, \nu}(Q^2)$, or the analytic analogs $\A_{\nu}(Q^2)$
of $a_{\rm pt}(Q^2)^{\nu}$. Furthermore, we construct the
timelike counterparts $\tH_{\nu}(\sigma)$ and $\ntH_{\nu}(\sigma)$
of the spacelike couplings $\tA_{\nu}(Q^2)$ and $\A_{\nu}(Q^2)$.
In sec.~\ref{sec:appl} we apply, as an illustration, the
presented method to evaluation of a timelike quantity,
the width of the Higgs decay into $b {\bar b}$ pair, $\Gamma(H \to b{\bar b})$.
The corresponding spacelike quantity has a perturbation expansion
which involves noninteger powers of $a_{\rm pt}$, due to the $b$-quark mass
anomalous dimension. 
We present the results for $\Gamma(H \to b{\bar b})$ 
as a function of the squared Higgs mass $s = M_H^2$, for various
analytic QCD scenarios and in pQCD. In sec.~\ref{sec:concl}
we present conclusions.

\section{Logarithmic noninteger derivatives of Euclidean coupling in any analytic QCD model}
\label{sec:lognonder}

As mentioned in the Introduction, we will start with the
analytic Euclidean coupling $\A_1(Q^2)$ and its logarithmic derivatives 
in a general
analytic QCD model. Such a model is determined (characterized)
fully by the discontinuity function
\be
\rho_1(\sigma) = {\rm Im} \ \A_1(Q^2=-\sigma - i \varepsilon) \ ,
\label{rho1}
\ee
defined for $\sigma \geq 0$. Usually, the discontinuity cut is
nonzero below a threshold value $-\sigma \leq -M_{\rm thr}^2$ where
$M_{\rm thr} \sim M_{\pi}$. The application of the Cauchy theorem
to the function $\A_1(Q^{' 2})/(Q^{'2} - Q^2)$ in the complex
$Q^{'2}$-plane, along the closed path made of a very large
circle and two segments just below and above the cut, gives us
the well known dispersion relation for the analytic Euclidean coupling $\A_1$
\be
\A_1(Q^2) = \frac{1}{\pi} \int_{0}^{+\infty} \ d \sigma 
\frac{ \rho_1(\sigma) }{(\sigma + Q^2)} \ ,
\label{disp}
\ee
where $q^2 \equiv - Q^2$ is non-Minkowskian, i.e., $Q^2$ can have any value in the complex plane except the cut $(-\infty, -M^2_{\rm thr}]$.
The logarithmic derivatives are defined as
\be
\tA_{n+1}(Q^2) \equiv \frac{(-1)^n}{\beta_0^n n!}
\frac{ \partial^n \A_1(Q^2)}{\partial (\ln Q^2)^n} \ ,
\qquad (n=0,1,2,\ldots) \ ,
\label{tAn}
\ee
where $\beta_0$ is the first coefficient of the $\beta$ function:
$\beta_0= (1/4)(11 - 2 n_f/3)$; 
$Q^2 d a_{\rm pt}/d Q^2 = - \beta_0 a_{\rm pt}^2 + {\cal O}(a_{\rm pt}^3)$. We note that
for $n=0$ equation (\ref{tAn}) gives $\tA_1 \equiv \A_1$.
We can write the
logarithmic derivatives in the following form:
\be
\tA_{n+1}(Q^2) = \frac{1}{\pi}
\int_{0}^{\infty} \ \frac{d \sigma}{\sigma} \rho_1(\sigma) 
\frac{1}{\beta_0^n \Gamma(n+1)} \frac{d^{n}}{d (\ln z)^n} 
\left( \frac{z}{1 + z} \right) {\Big |}_{z=\sigma/Q^2} \ .
\label{disptAn1}
\ee
It turns out that the integrand is the known polylogarithm function
\be
 \frac{d^n}{d (\ln z)^n} \left( \frac{z}{1 + z} \right)
= 
\left(z\frac{d}{dz}\right)^n \, \sum_{m=1}^{\infty} (-1)^{m+1} z^m 
=  \sum_{m=1}^{\infty} (-1)^{m+1} m^n  z^m
= (-1) {\rm Li}_{-n}( -z) \ ,
\label{Li}
\ee
which brings equation (\ref{disptAn1}) in the following form:
\be
\tA_{n+1}(Q^2) = \frac{1}{\pi} \frac{(-1)}{\beta_0^n \Gamma(n+1)}
\int_{0}^{\infty} \ \frac{d \sigma}{\sigma} \rho_1(\sigma)  
{\rm Li}_{-n} ( -\sigma/Q^2 ) \ .
\label{disptAn2}
\ee
This relation is valid for $n=0,1,2,...$. Analytic continuation
in $n \mapsto \nu$ gives us\footnote{
In Mathematica \cite{Math8}, the ${\rm Li}_{-\nu}(z)$ function
is implemented as ${\rm PolyLog}[-\nu,z]$. However, at large $|z| > 10^7$,
${\rm PolyLog}[-\nu,z]$ appears to be unstable. For such $z$ we should 
use the identities relating ${\rm Li}_{-\nu}(z)$ with ${\rm Li}_{-\nu}(1/z)$, 
which can be found, for example, in \cite{Erdelyi}.} 
the logarithmic noninteger derivatives
\be
\tA_{\nu+1}(Q^2) = \frac{1}{\pi} \frac{(-1)}{\beta_0^{\nu} \Gamma(\nu+1)}
\int_{0}^{\infty} \ \frac{d \sigma}{\sigma} \rho_1(\sigma)  
{\rm Li}_{-\nu}\left( - \frac{\sigma}{Q^2} \right) \quad (-1 < \nu) \ .
\label{disptAn3}
\ee
We note that the integral converges for $\nu > -1$. Namely, at high
$\sigma$ ($|z| \gg 1$ where $z \equiv \sigma/Q^2$) we have in the integrand of
equation (\ref{disptAn3}): $\rho_1(\sigma) \approx
\rho_1^{\rm (pt)}(\sigma) \sim \ln^{-2} \sigma \sim \ln^{-2} z$ and 
${\rm Li}_{-\nu}(-z) \sim \ln^{-\nu} z$ (for noninteger $\nu$). 
Therefore, the integral converges at 
$\sigma \to \infty$ if $\nu > -1$. The integral obviously converges at low $\sigma$, too.

In principle, a continuation to arbitrary $\nu$, as performed by
the transition from equation (\ref{disptAn2}) to equation (\ref{disptAn3}), could 
in principle miss some terms, such as terms proportional to 
$\sin^k ( \nu \pi)$. However, such terms will be
excluded because they are finite oscillatory when $\nu \to \pm \infty$.

It is interesting that the recursive relation 
\begin{equation}
\tA_{\nu+2}(Q^2) = \frac{(-1)}{\beta_0 (\nu+1)} 
\frac{d}{d \ln Q^2} {\tA}_{\nu+1}(Q^2) \ ,
\label{tArecur}
\end{equation}
which for positive integer $\nu=n=0,1,2,\ldots$ is a direct consequence of
the definition (\ref{disptAn1}), remains valid even for noninteger $\nu$ as
a consequence of the relation (\ref{disptAn3}) and the known\footnote{
This relation can be obtained, for example, by applying $d/d \ln z$ 
to the power series 
$Li_{\nu^{\prime}}(z)_{\rm ser.} = \sum_{m=1}^{\infty} z^m m^{-\nu^{\prime}}$. } 
relation $z (d/dz) {\rm Li}_{-\nu}(z) = {\rm Li}_{-\nu-1}(z)$.

We can recast the result (\ref{disptAn3}) into an alternative form involving
the spacelike coupling $\A_1$ instead of the discontinuity function
$\rho_1(\sigma)$. This can be performed in the following way.

We can use the following integral form of ${\rm Li}_{-\nu}$ 
function (\cite{PBM})\footnote{
Equation (\ref{Li_nu}) can be proven by expanding the integrand in powers of
$e^{-t}$ and using the basic integral expression for the $\Gamma(\nu^{'})$ function
(where $\nu^{'} \equiv  - \nu > 0$): $\int_0^{\infty} du \; e^{-u} u^{\nu^{'} - 1} = \Gamma(\nu^{'})$.
In this way, the (convergent for $|z|<1$) series $\sum_{m=1}^{\infty} z^m m^{-\nu^{'}}$
is generated, which is just the polylogarithm function ${\rm Li}_{\nu^{'}}(z)
\equiv {\rm Li}_{-\nu}(z)$.}
appearing in equation (\ref{disptAn3}):
\be
{\rm Li}_{-\nu}(z) = \frac{z}{\Gamma(-\nu)} \int_0^{\infty} 
\frac{ dt \; t^{-\nu -1} }{(e^t - z)} = 
 \frac{z}{\Gamma(-\nu)} \int_0^1 \frac{d \xi}{1 - z \xi} 
\ln^{-\nu -1} \left( \frac{1}{\xi} \right) \quad (\nu < 0) \ .
\label{Li_nu}
\ee
The last expression on the right-hand side
was obtained by the change of variable $t = \ln(1/\xi)$. 
Since we have in our result (\ref{disptAn3}) ${\rm Li}_{-\nu}$ with
$-1 < \nu$ (and not just: $-1 < \nu < 0$), we extend the integral representation
to higher $\nu > 0$. This is achieved by using in equation (\ref{Li_nu}) 
the aforementioned relation  $(d/d \ln z) {\rm Li}_{-\nu} = {\rm Li}_{-\nu - 1}$.
We thus obtain, for $\nu = n + \delta$, with $0 < \delta < 1$ and $n=-1, 0, 1, 2, \ldots$,
the following integral form, \cite{Kotikov:2010bm}:
\be
{\rm Li}_{-n-\delta}(z) = \left( \frac{d}{d \ln z} \right)^{n+1}
\left[ \frac{z}{\Gamma(1 - \delta) } \int_0^1 \frac{d \xi}{1 - z \xi}
\ln^{-\delta} \left( \frac{1}{\xi} \right) \right] \quad
(n=-1, 0, 1, \ldots; \ 0 < \delta < 1) \ .
\label{Li_nugen}
\ee
Inserting the representation (\ref{Li_nugen}), for $\nu = n + \delta$, into
our general formula (\ref{disptAn3}), and exchanging the order of integration,
gives us
\bea
\tA_{\nu+1}(Q^2) &=& 
\frac{1}{\beta_0^{\nu} \Gamma(n+1+\delta)\Gamma(1-\delta)} 
\left(- \frac{d}{d \ln Q^2}\right)^{n+1}
\int_0^1 \frac{d \xi}{\xi} \ln^{-\delta}\left(\frac{1}{\xi}\right)
\int_{0}^{\infty} 
\frac{d \sigma \rho_1(\sigma)}{\pi (\sigma + Q^2/\xi)} \ .
\label{disptAn4}
\eea
The last integral over $d \sigma$ is the spacelike coupling $\A_1(Q^2/\xi)$ due to
the dispersion relation (\ref{disp}). Therefore, we obtain the alternative form
of the result (\ref{disptAn3}), for  $\nu = n + \delta$, with $0 < \delta < 1$ and 
$n=-1, 0, 1, 2, \ldots$,
\bea
\tA_{\nu+1}(Q^2) \equiv \tA_{n+1+\delta}(Q^2) &=& 
\frac{1}{\beta_0^{\nu} \Gamma(1+\nu)\Gamma(1-\delta)} 
\left(- \frac{d}{d \ln Q^2}\right)^{n+1}
\int_0^1 \frac{d \xi}{\xi} \A_1(Q^2/\xi) \ln^{-\delta}\left(\frac{1}{\xi}\right) 
\label{disptAn5a}
\\
& = &
\frac{1}{\beta_0^{\nu}} \frac{\Gamma(1+\delta)}{\Gamma(n+1+\delta)}
\frac{ \sin(\pi \delta) }{(\pi \delta)}
\left(- \frac{d}{d \ln Q^2}\right)^{n+1}
\int_0^{\infty} \frac{dt}{t^{\delta}} \A_1(Q^2 e^t) \ ,
\label{disptAn5b}
\eea
where the last form (\ref{disptAn5b}) was obtained from the previous one by
the substitution $t = \ln(1/\xi)$ and using the identity 
$\Gamma(1 + \delta) \Gamma(1 - \delta) = \pi \delta/\sin(\pi \delta)$.

Furthermore, we will now prove that the obtained result (\ref{disptAn3})
(equivalent to equations (\ref{disptAn5a}) and (\ref{disptAn5b})) reproduces,
in the specific case of the Minimal Analytic model (MA) of \cite{ShS,MSS,Sh} 
at one-loop level, the explicit
result obtained in \cite{BMS1}\footnote{
Note that $a$ in \cite{BMS1,BMS2,BMS3,BP} corresponds to our $\beta_0 a$
(with our $\beta_0 = (1/4)(11 - 2 n_f/3)$); their $\A_{\nu+1}$ corresponds to
our $\beta_0^{\nu+1} \A_{\nu + 1}$; they use the transcendental Lerch
function notation $z \; \Phi(z,\nu^{'},1) \equiv F(z, \nu^{'})$ 
for the polylogarithm function ${\rm Li}_{\nu^{'}}(z)$. On the other hand,
$\A_n$ in \cite{ShS,MSS,Sh} corresponds to analytic analogs
of $\alpha_s^n = \pi^n a^n$, i.e., their $\A_n$ corresponds to
our $\pi^n \A_n$.}

\be
\tA_{\nu+1}(Q^2)^{\rm (MA, 1-\ell)}
= \A_{\nu+1}(Q^2)^{\rm (MA, 1-\ell)} = \frac{1}{\beta_0^{\nu+1}}
\left(  \frac{1}{\ln^{\nu+1}(Q^2/\bL^2)} -
\frac{ {\rm Li}_{-\nu}(\bL^2/Q^2)}{\Gamma(\nu+1)} \right) \ ,
\label{disptAn31l}
\ee
where the scale $\bL$ appears in the
one-loop MA analytic coupling $ \A_1(Q^2)^{\rm (MA, 1-\ell)}$ and in its
discontinuity function $\rho_1(\sigma)_{\rm pt}^{\rm (1-\ell)}$
\bea
\A_1(Q^2)^{\rm (MA, 1-\ell)} & = & \frac{1}{\beta_0}
\left(  \frac{1}{\ln(Q^2/\bL^2)} - \frac {\bL^2}{(Q^2 - \bL^2)} \right) \ ,
\label{A11l}
\\
\rho_1(\sigma)_{\rm pt}^{\rm (1-\ell)} &=& 
{\rm Im} a_{\rm pt}(-\sigma - i \epsilon)^{\rm (1-\ell)} = 
{\rm Im} \A_1(-\sigma - i \epsilon)^{\rm (MA, 1-\ell)}
\nonumber\\
&=&  \frac{1}{\beta_0} 
{\rm Im} \frac{1}{ \left( \ln(\sigma/\bL) - i \pi \right) }
= \frac{\pi}{\beta_0} \frac{1}{\left( \ln^2(\sigma/\bL) + \pi^2 \right)} \ .
\label{rho11l}
\eea
When replacing $\A_1(Q^2/\xi)$ in the integrand of the expression (\ref{disptAn5a})
by the second term of the expression (\ref{A11l}) for $\A_1(Q^2/\xi)^{\rm (MA, 1-\ell)}$,
and using the integral form (\ref{Li_nugen}) for ${\rm Li}_{-\nu}$, we obtain
immediately
\bea
\frac{1}{\beta_0^{\nu} \Gamma(1+\nu)\Gamma(1-\delta)} 
\left(- \frac{d}{d \ln Q^2}\right)^{n+1}
\int_0^1 \frac{d \xi}{\xi} \ln^{-\delta}\left(\frac{1}{\xi}\right)
\frac{1}{\beta_0}\frac{(-1)\bL^2}{(Q^2/\xi - \bL^2)} & = &
\frac{(-1)}{\beta_0^{\nu+1} \Gamma(\nu+1)}  {\rm Li}_{-\nu}(\bL^2/Q^2) \ .
\label{1l2ndterm}
\eea
On the other hand, when replacing $\A_1(Q^2/\xi)$ in the integrand of the expression 
(\ref{disptAn5b}) by the first term of the expression (\ref{A11l}) 
for $\A_1(Q^2/\xi)^{\rm (MA, 1-\ell)}$, we obtain in a direct manner\footnote{
\label{intdy}
We can use the integration variable $y=t/t_0$, where $t_0=\ln(Q^2/\bL^2)$, and
the exact solution of the following integral:
\begin{displaymath}
\int_0^{\infty} \frac{d y}{y^{\delta} (y + 1)} = \frac{\pi}{\sin(\pi \delta)} \ ,
\qquad {\rm where \ } 0 < \delta < 1 \ .
\end{displaymath}
}
\bea
\frac{1}{\beta_0^{\nu}} \frac{\Gamma(1+\delta)}{\Gamma(n+1+\delta)}
\frac{ \sin(\pi \delta) }{(\pi \delta)}
\left(- \frac{d}{d \ln Q^2}\right)^{n+1}
\int_0^{\infty} \frac{dt}{t^{\delta}} \frac{1}{\beta_0} 
\frac{1}{\left[ t + \ln(Q^2/\bL^2) \right]}
& = & \frac{1}{\beta_0^{\nu+1} \ln^{\nu+1}(Q^2/\bL^2)} \ .
\label{1l1stterm}
\eea
Combining the results (\ref{1l2ndterm}) and (\ref{1l1stterm}), we obtain the
full result (\ref{disptAn31l}) for $\A_{\nu+1}(Q^2)$ in the 
one-loop approach of MA, for any noninteger $\nu$ such that $-1 < \nu$ (when
$\nu$ is nonnegative integer, the limit $\delta \to 0$ can be made in the
derivation). This (one-loop MA) result, obtained for the first time
by Bakulev, Mikhailov and Stefanis (BMS) in \cite{BMS1}, 
has several interesting properties, 
as pointed out in \cite{BMS2} (their equations (3.14)-(3.19)).
The result  (\ref{disptAn31l}) is explicit and allows us to apply it even for
$\nu \leq -1$, and even for complex $\nu$; 
this is a kind of analytic continuation in $\nu$. We can thus use this
result, by adding and subtracting it from of our general integral 
expression (\ref{disptAn3}), thus extending the $\nu$-regime of applicability
of our expression
\bea
\tA_{\nu+1}(Q^2) &=& \tA_{\nu+1}(Q^2)^{\rm (MA,1-\ell)} +
\frac{1}{\pi} \frac{(-1)}{\beta_0^{\nu} \Gamma(\nu+1)}
\int_{0}^{\infty} \ \frac{d \sigma} {\sigma} 
\left[ \rho_1(\sigma) - \rho_1(\sigma)_{\rm pt}^{\rm (1-\ell)} \right]   
{\rm Li}_{-\nu}\left( - \frac{\sigma}{Q^2} \right) \quad (-2 < \nu) \ ,
\label{disptAn3b}
\eea
where $\tA_{\nu+1}(Q^2)^{\rm (MA,1-\ell)}$ and $\rho_1(\sigma)_{\rm pt}^{\rm (1-\ell)}$
are given in equations (\ref{disptAn31l}) and (\ref{rho11l}), respectively.
Now the integral converges also for $-2 < \nu < -1$, because, 
due to asymptotic freedom,
the difference $[ \rho_1(\sigma) - \rho_1(\sigma)_{\rm pt}^{\rm (1-\ell)}]$ behaves at
large $\sigma$ as $\sim \ln \ln \sigma/\ln^3 \sigma$ and not as $1/\ln^2 \sigma$.
Further, the expression (\ref{disptAn3b}) implies that 
$\tA_{0}(Q^2)$ [$\equiv {\rm lim}_{\nu \to -1} \tA_{\nu+1}(Q^2)$] $=1$ for all
complex $Q^2$, because: ${\rm Li}_{-\nu}(z)/ \Gamma(\nu+1) \to 0$ when
$\nu \to -1$, and $\tA_{0}(Q^2)^{\rm (MA,1-\ell)} \equiv 1$.

\section{Analytization procedure for observables with noninteger 
powers of coupling}
\label{sec:man}
 
In QCD we encounter often spacelike (Euclidean) observables ${\cal F}(Q^2)$
whose perturbative expansion starts with a noninteger power
$a_{\rm pt}^{\nu_0}$
\be
{\cal F}(Q^2)_{\rm pt} = a_{\rm pt}(Q^2)^{\nu_0} + {\cal F}_1 a_{\rm pt}(Q^2)^{\nu_0+1} +
 {\cal F}_2 a_{\rm pt}(Q^2)^{\nu_0+2} + \cdots
\label{calFpt}
\ee
The general analytization procedure of the pQCD-evaluated
observables with integer powers is
\be
{\ta}_{{\rm pt},n+1} \mapsto \tA_{n+1}
\quad (n=0,1,2,\ldots) \ ,
\label{anrule1}
\ee
where ${\ta}_{{\rm pt},n+1}$ are the logarithmic derivatives
of the pQCD coupling $a_{\rm pt}$
\be
{\ta}_{{\rm pt},n+1}(Q^2)
\equiv \frac{(-1)^n}{\beta_0^n n!}
\frac{ \partial^n a_{\rm pt}(Q^2)}{\partial (\ln Q^2)^n} 
= a_{\rm pt}^n + {\cal O}(a_{\rm pt}^{n+1})
\qquad (n=0,1,2,\ldots) \ ,
\label{tan}
\ee
and\footnote{
Naively, one might suppose that the analytization procedure, in the
evaluation of observables ${\cal F}(Q^2) \equiv {\cal D}(Q^2)$ 
with integer powers of $a_{\rm pt}$,
in any given anQCD model would be 
$a_{\rm pt}^{n+1} \mapsto A_1^{n+1}$. It turns out that, in those anQCD models
whose $\A_1(Q^2)$ at high $Q^2$ differs from $a_{\rm pt}(Q^2)$ by
negative powers of $Q^2$ ($\sim (\Lambda^2/Q^2)^k$), such naive
analytization procedure leads to strong renormalization
scheme (RS) dependence of the truncated (modified) analytic series 
${\cal D}^{(N)}(Q^2)_{\rm (m)an}$,
due to the contributions of power terms $\sim (\Lambda^2/Q^2)^m$ to the
derivative $\partial {\cal D}^{(N)}(Q^2)_{\rm (m)an}/\partial {\rm RS}$,
see \cite{CV2}.}
$\beta_0$ is the first coefficient of the $\beta$-function
\bea
\frac{d a_{\rm pt}(\mu^2)}{d \ln \mu^2} \equiv
\beta(a_{\rm pt}) & =& -\beta_0 a_{\rm pt}^2 -\beta_1 a_{\rm pt}^3 -\beta_2 a_{\rm pt}^4 -\beta_3 a_{\rm pt}^5 -\beta_4 a_{\rm pt}^6
 - \dots
\nonumber \\
 &=& -\beta_0 a_{\rm pt}^2 \, \left(1+ c_1 a_{\rm pt} +c_2 a_{\rm pt}^2 +c_3 a_{\rm pt}^3 +c_4 a_{\rm pt}^4 + \dots \right)
~~~\left(c_j = \frac{\beta_j}{\beta_0} \right) \ .
\label{RGE}
\eea
In the case of observables whose pQCD-evaluated expressions
are the (truncated) expansions equation (\ref{calFpt}) with
noninteger $\nu_0$, the analytization procedure (\ref{anrule1})
is naturally extended to (\cite{CV1,CV2}) 
\be
{\ta}_{{\rm pt},\nu+1} \mapsto \tA_{\nu+1} \ ,
\label{anrule2}
\ee
where the expression for $\tA_{\nu+1}$ is given in equation (\ref{disptAn3}).
Therefore, at this stage, the problem of evaluation of such observables
in anQCD is reduced to re-expressing the noninteger powers
$a_{\rm pt}^{\nu}$ in pQCD expansion (\ref{calFpt})
in terms of the logarithmic noninteger derivatives
${\ta}_{{\rm pt},\nu+m}(Q^2)$, in order to perform the
subsequent analytization via equation (\ref{anrule2}). Stated otherwise,
we find first the coefficients $k_m(\nu)$ of the relations
\be
{\ta}_{{\rm pt},\nu} ~=~ a_{\rm pt}^{\nu} + \sum_{m=1}^{\infty}
k_m(\nu) a_{\rm pt}^{\nu + m} \ ,
\label{tanuanu}
\ee
and, as a consequence, the coefficients $\tk_m(\nu)$ of the inverse relations
\be
a_{\rm pt}^{\nu} ~=~ {\ta}_{{\rm pt},\nu} + \sum_{m=1}^{\infty}
\tk_m(\nu) {\ta}_{{\rm pt},\nu + m} \ .
\label{anutanu}
\ee
The expressions for the coefficients $k_m(\nu)$ (and $\tk_m(\nu)$) are derived
in appendix \ref{App1}; see equations (\ref{kmsk1})-(\ref{kmsk4}),
(\ref{Baux}) and (\ref{tkms}) there for explicit expressions.
 There, the coefficients $k_m(n)$ and $\tk_m(n)$, for $n$ integer, are obtained
by solving the difference (recursion) equations relating $k_m(n+1)$ with
$k_m(n)$, $k_{m-1}(n)$, etc. The solution for $k_m(n)$ (and $\tk_m(n)$) 
is obtained
in a form involving combinations of Gamma functions  $\Gamma(x)$ and 
their derivatives (up to $m$ derivatives), 
at the values of the argument $x=1$ and $x=+n+m'$ 
(for $m'=1,\ldots,m$).
In the obtained expressions, the integer $n$ is then replaced 
by an arbitrary noninteger $\nu$
($n \mapsto \nu$). The latter step is an analytic continuation 
similar to the step $n \mapsto \nu$
from equation (\ref{disptAn2}) to equation (\ref{disptAn3}).

The relations (\ref{anutanu}) allow us to reexpress the expansion (\ref{calFpt}) in terms of ${\ta}_{{\rm pt},\nu}$'s 
\be
{\cal F}(Q^2)_{\rm mpt} = {\ta}_{\rm pt, \nu_0}(Q^2) + 
{\widetilde {\cal F}}_1 {\ta}_{\rm pt,\nu_0+1}(Q^2) + 
{\widetilde {\cal F}}_2 {\ta}_{\rm pt,\nu_0+2}(Q^2) + \cdots \ ,
\label{Fmpta}
\ee
where 'mpt' stands for ``modified perturbation series'' and
the coefficients ${\widetilde {\cal F}}_n$ are related with the
coefficients ${\cal F}_n$ of the original perturbation series (\ref{calFpt})
via relations involving the coefficients $\tk_m(\nu_0+n)$ appearing in the
relations (\ref{anutanu}) 
\be
{\widetilde {\cal F}}_1 = {\cal F}_1 + \tk_1(\nu_0), \quad
{\widetilde {\cal F}}_2 = {\cal F}_2 + {\cal F}_1 \tk_1(\nu_0+1) + \tk_2(\nu_0) \ , \quad {\rm etc.}
\label{tFjFj}
\ee
See equations (\ref{tF1})-(\ref{tF4}) in appendix \ref{App1} for 
more relations involving higher orders. 

At this stage we apply the analytization procedure
(\ref{anrule2}) to obtain the ``modified analytic'' (man) series\footnote{
The word ``modified'' is used here because
we are analytizing the
logarithmic (noninteger) derivatives of $a_{\rm pt}$ [equation (\ref{tan})]
and not the (noninteger) powers of $a_{\rm pt}$. Our method
leading to the expression (\ref{disptAn3}) makes the described 
former ``modified'' analytization approach more direct than the latter
(equivalent) analytization approach involving (noninteger) powers
 [cf.~equation (\ref{anrule3}) later in this section.]} 
for the (dimensionless) spacelike quantity ${\cal F}(Q^2)$
\be
{\cal F}(Q^2)_{\rm man} = {\tA}_{\nu_0}(Q^2) + 
{\widetilde {\cal F}}_1 {\tA}_{\nu_0+1}(Q^2) + 
{\widetilde {\cal F}}_2 {\tA}_{\nu_0+2}(Q^2) + \cdots \ ,
\label{Fmana}
\ee
where the expressions for ${\tA}_{\nu_0+1}(Q^2)$, ${\tA}_{\nu_0+2}(Q^2)$ are
given, in any given anQCD, by equation (\ref{disptAn3}).\footnote{
Similarly as was argued in \cite{CV2} in the case when
$\nu_0$ is integer ($\nu_0=1$), it can be shown that the truncated series 
${\cal F}(Q^2;\mu^2)^{[N]}_{\rm man}$,
whose last included term is $\sim \tA_{\nu_0+N}(\mu^2)$ 
and the renormalization scale $\mu^2$ is used, has a systematically 
suppressed renormalization scale 
dependence when the order index $N$ increases
$\partial {\cal F}(Q^2;\mu^2)^{[N]}_{\rm man}/\partial \ln \mu^2 =
{\cal O}({\tA}_{\nu_0+N+1})$. This is really a systematic suppression,
because in analytic QCD models we have the hierarchy
$|\tA_{\nu_0}(\mu^2)| > |\tA_{\nu_0+1}(\mu^2)| > \cdots$, for all
complex $\mu^2$ outside the cut.}

\begin{figure}[htb] 
\centering\epsfig{file=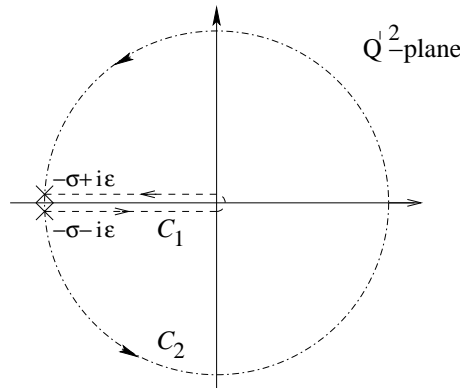,width=6.cm}
\vspace{-0.4cm}
 \caption{\footnotesize  Paths ${\cal C}_1$ and ${\cal C}_2$
in the complex $Q^{' 2}$-plane.}
\label{contour12}
 \end{figure}

On the other hand, an observable ${\cal T}(\sigma)$ that is related with
a spacelike observable ${\cal F}(Q^2)$ via the integral transformation
\begin{equation}
{\cal F}(Q^2) = Q^2 \int_0^{\infty} 
\frac{d \sigma \ {\cal T}(\sigma)}{(\sigma + Q^2)^2}
\label{FT}
\end{equation}
is timelike (Minkowskian). The inverse transformation is
\begin{equation}
{\cal T}(\sigma) = \frac{1}{2 \pi i} 
\int_{-\sigma - i \varepsilon}^{-\sigma + i \varepsilon} 
\frac{d Q^{' 2}}{Q^{' 2}} {\cal F}(Q^{' 2}) \ ,
\label{TF}
\end{equation}
where the integration contour is in the complex 
$Q^{' 2}$-plane encircling the singularities of the integrand,
e.g., path ${\cal C}_1$ or ${\cal C}_2$ of fig.~\ref{contour12}. 
Application of the tranformation  (\ref{TF})
to the (modified) analytic series (\ref{Fmana}) then gives for the
timelike quantity ${\cal T}(\sigma)$ the following (Minkowskian) 
``modified analytic series''
\be
{\cal T}(\sigma)_{\rm man} = {\tH}_{\nu_0}(\sigma) + 
{\widetilde {\cal F}}_1 {\tH}_{\nu_0+1}(\sigma) + 
{\widetilde {\cal F}}_2 {\tH}_{\nu_0+2}(\sigma) + \cdots \ ,
\label{calTman}
\ee 
where the timelike (Minkowskian) 
couplings ${\tH}_{\nu+1}(\sigma)$ are defined as
\begin{equation}
{\tH}_{\nu+1}(\sigma) \equiv \frac{1}{2 \pi i} 
\int_{-\sigma - i \varepsilon}^{-\sigma + i \varepsilon} 
\frac{d Q^{' 2}}{Q^{' 2}} {\tA}_{\nu+1}(Q^{' 2}) \ ,
\label{tHa}
\end{equation}
and the inverse transformation is
\be
\tA_{\nu+1}(Q^2) = Q^2 \int_0^{\infty} 
\frac{d \sigma \; {\tH}_{\nu+1}(\sigma)}{(\sigma + Q^2)^2} \ .
\label{tAtHA}
\ee

Here we would like to note that the reexpression of the expansion 
(\ref{calFpt}) to the one of equation (\ref{Fmpta}) 
is a well-defined operation for quite
convergent series. It is well known that 
mostly the QCD series are assumed to be asymptotic
(see, for example, \cite{Kazakov:1980rd}), and some arguments to 
justify such a reexpression should be 
done,\footnote{We thank to anonimous Referee who drew our attention to this 
possible problem.}
despite the fact that we use in our analysis only the first several terms in
the expansions (\ref{calFpt}) and (\ref{Fmpta}).
Some cases of different types of asymptotics were recently considered in
\cite{Shirkov:2012zm}.

To show the correctness of the reexpression and, at the same time, to avoid
an additional increase of the volume of the main text part of our paper, 
we consider in appendix \ref{App2} the standard 
Lipatov-type behavior \cite{Lipatov:1976ny} for the $n$th term of the 
expansion (\ref{calFpt}) and recover 
the similar behavior for the $n$th term of the 
expansion (\ref{Fmpta}). We show that there is
even a slight weakening of the rise for the 
$n$th term of the expansion (\ref{Fmpta}) in comparison with the original one
in the expansion (\ref{calFpt}) if the coefficients in (\ref{calFpt}) are
nonalternating in sign.\footnote{A decrease of 
${\widetilde {\cal F}}_n$ in comparison with
${\cal F}_n$ for $n=3$ and $4$ has been earlier observed also in 
\cite{Cvetic:2010ut}.} 

The parameter  ${\ta}_{\rm pt,\nu_0+n}(Q^2)$ of the expansion (\ref{Fmpta})
is very close to the corresponding one $a_{\rm pt}^{\nu_0+n}(Q^2)$ in the
high-momentum regime (i.e., when $a_{\rm pt}$-values are small).
This observation is consistent with the similarity of the coefficients 
${\cal F}_n$ and ${\widetilde {\cal F}}_n$ shown in appendix \ref{App2},
and another argument for acceptability of the reexpression of the expansion 
(\ref{calFpt}) to the one of equation (\ref{Fmpta}).
  
We note also that, after the analytization procedure (\ref{anrule2}), there is 
in general a significant supppression of the
new parameters $\tA_{\nu_0+n}(Q^2)$ in comparison 
with ${\ta}_{\rm pt,\nu_0+n}(Q^2)$ and $a_{\rm pt}^{\nu_0+n}(Q^2)$
(see a recent review \cite{Bakulev} and discussions therein).

\subsection{Timelike (Minkowskian) coupling parameter}
\label{subsec:timelike}

\begin{figure}[htb] 
\centering\epsfig{file=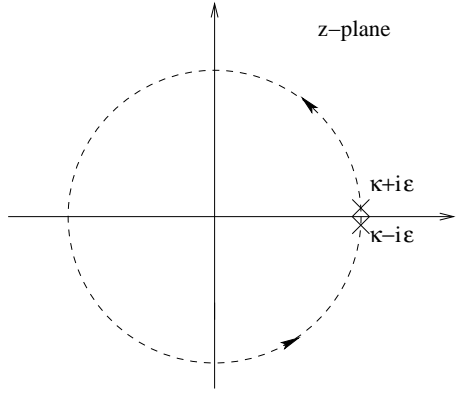,width=60mm}
\vspace{-0.4cm}
\caption{Integration contour for the integral (\ref{contourid}).}
\label{contourz}
 \end{figure}

Direct use of the expression (\ref{disptAn3}) in the integral (\ref{tHa})
gives us a double integral. When $-1 < \nu < 1$, the order of integration 
can be exchanged because the resulting double integral is convergent, 
and we obtain
\begin{equation}
{\tH}_{\nu+1}(\sigma) = \frac{(-1)}{2 \pi i \beta_0^{\nu} \Gamma(\nu+1)}
\int_0^{\infty} \frac{d \sigma^{\prime}}{\sigma^{\prime}} 
\rho_1(\sigma^{\prime})
\int_{-\sigma - i \varepsilon}^{-\sigma + i \varepsilon} 
\frac{d Q^{' 2}}{Q^{' 2}} {\rm Li}_{-\nu}(-\sigma^{\prime}/Q^{' 2}) \ .
\label{tlAb}
\end{equation}
Now we can use the fact that ${\rm Li}_{-\nu}(z)$ is a function with a 
branch cut discontinuity $[1,+\infty)$ in the complex $z$-plane, and 
the earlier mentioned relation
$ {\rm Li}_{-\nu}(z) = d {\rm Li}_{1-\nu}(z)/ d \ln z$, to obtain the identity
\begin{equation}
\frac{1}{2 \pi i} \oint_{|z|=\kappa, {\rm pos.dir.}} 
\frac{dz}{z} {\rm Li}_{-\nu}(z) 
= \frac{1}{\pi} {\rm Im} {\rm Li}_{1 - \nu} (\kappa - i \varepsilon) 
= - \frac{1}{\Gamma(1-\nu)} (\ln \kappa)^{-\nu} \Theta(\kappa - 1) \ ,
\label{contourid}
\end{equation}
where $\Theta$ on the right-hand side is the Heaviside step function,
and the integration on the left-hand side is along the 
contour of radius $|z|=\kappa$ over the angles $\Phi \equiv {\rm arg}(z)$ from 
$+0$ to $(2 \pi - 0)$, see fig.~\ref{contourz}.  
Using the relation 
(\ref{contourid}) in the integration over $Q^{' 2}$ on the right-hand 
side of equation (\ref{tlAb}), we obtain the simplified
expression for the general Minkowskian coupling
${\tH}_{\nu+1}(\sigma)$ in any analytic QCD and for any real $\nu$ in 
the interval $-1 < \nu < 1$
\begin{equation}
{\tH}_{\nu+1}(\sigma) = \frac{\sin (\pi \nu)}{\pi^2 \nu \beta_0^{\nu}}
\int_0^{\infty} \frac{d w}{w^{\nu}} \rho_1(\sigma e^w) \ ,
\quad (-1 < \nu < 1) \ ,
\label{tHnu1}
\end{equation}
and we used here a new variable $w=\ln(\sigma^{\prime}/\sigma)$. 
The integral (\ref{tHnu1}) is clearly convergent
at $w \to 0$. It is also convergent at $w \to + \infty$, because there
$\rho_1(\sigma e^w) \approx \rho_1^{\rm (pt)}(\sigma e^w) \sim 1/w^2$.

The case of
${\ntH}_1(\sigma)$ is obtained in the limit $\nu \to +0$ of the above expression
\begin{equation}
{\ntH}_1(\sigma) = {\tH}_{1}(\sigma) = \frac{1}{\pi}
\int_0^{\infty} d w \ \rho_1(\sigma e^w) =
\frac{1}{\pi}
\int_{\sigma}^{\infty}
 \frac{d \sigma^{\prime}}{\sigma^{\prime}} \rho_1(\sigma^{\prime})
\ ,
\label{tH1}
\end{equation}
which is a well known result. 
For Minkowskian couplings 
${\tH}_{\nu+1} = {\tH}_{\delta+1+n}(\sigma)$ with higher index $\nu=n+\delta$ 
($n=0,1,2,\ldots$; and $0 < \delta < 1$),
we can use the recursion formulas
\begin{equation}
{\tH}_{\nu+2}(\sigma) = 
\frac{(-1)}{\beta_0 (\nu+1)} 
\frac{d}{d \ln \sigma} {\tH}_{\nu+1}(\sigma) \ ,
\label{tHrecur}
\end{equation}
which can be obtained from the relations (\ref{tArecur}) and (\ref{tHa}),
and obtain (see appendix \ref{App3} for derivation)
\begin{eqnarray}
{\tH}_{\nu+1}(\sigma) & = & 
\frac{\sin(\pi(\delta+n))}{\pi^2 (\delta+n) \beta_0^{\delta+n}}
\int_0^{\infty} \frac{d w}{w^{\delta+n}} 
\left[
\rho_1(\sigma e^w) - \rho_1(\sigma) 
- \frac{w}{1!} \frac{d \rho_1(\sigma)}{d \ln \sigma}
 - \ldots
-  \frac{w^{n-1}}{(n-1)!} \frac{d^{n-1} \rho_1(\sigma)}{d (\ln \sigma)^{n-1}} \right] \ ,
\label{tHnun1}
\end{eqnarray}
where $\nu=n+\delta$, with $0 < \delta < 1$ and $n=0,1,2,\ldots$. When $n=0$,
the expression in the brackets in equation (\ref{tHnun1}) is 
$[\rho_1(\sigma e^w)]$, and $\delta= \nu$ in this case varies in a larger interval 
$-1 < \delta < 1$, i.e., equation (\ref{tHnu1}).
The integral (\ref{tHnun1}) is clearly convergent at 
$w \to + \infty$. It is also convergent at $w \to 0$, because the expression in backets
behaves as $\sim w^n$ there.

The version of this formula when $\delta=0$ is obtained by repeated application
of the recursion formula (\ref{tHrecur}) to the expression (\ref{tH1})
\be
{\tH}_{n+1}(\sigma) = \frac{(-1)^n}{\beta_0^n n!}
\frac{ \partial^n \ntH_1(\sigma)}{\partial (\ln \sigma)^n} 
= \frac{(-1)^{n-1}}{\beta_0^n n!} \frac{1}{\pi} 
\frac{d^{n-1} \rho_1(\sigma)}{d (\ln \sigma)^{n-1}} \ ,
\qquad (n=1,2,\ldots) \ .
\label{tHn1}
\ee

As we did in the previous section for the spacelike 
coupling $\tA_{\nu+1}$, we derive now from our general 
timelike coupling (\ref{tHnun1}) the explicit result obtained in 
\cite{BMS2} for the one-loop MA case. First we rewrite the
integrand in equation (\ref{tHnun1})
\be
\rho_1(\sigma e^w) - \sum_{k=0}^{n-1} \frac{w^k}{k!}
\frac{d^k \rho_1(\sigma)}{d(\ln \sigma)^k}
~=~  \sum_{k=n}^{\infty} \frac{w^k}{k!}
\frac{d^k \rho_1(\sigma)}{d(\ln \sigma)^k}
\ .
\label{dop2.1}
\ee
Using the one-loop MA expression for $\rho_1(\sigma)_{\rm pt}^{\rm (1-\ell)}$, 
equation (\ref{rho11l}), we can represent (\ref{dop2.1}) in this case as
\bea
\sum_{k=n}^{\infty} \frac{w^k}{k!}
\frac{d^k \rho_1(\sigma)_{\rm pt}^{\rm (1-\ell)}}{d(\ln \sigma)^k}
& = & \frac{1}{\beta_0} \; 
{\rm Im} 
\frac{(-w)^n}{(\ln (\sigma/\bL^2) -i\pi)^{n+1}}
 \sum_{\ell=0}^{\infty} \frac{(-w)^\ell}{(\ln (\sigma/\bL^2) -i\pi)^\ell}
\nonumber\\
&=& \frac{1}{\beta_0} \; {\rm Im} \frac{(-w)^n}{(\ln (\sigma/\bL^2) -i\pi)^n} \;
\frac{1}{(\ln (\sigma/\bL^2) +w -i\pi)}
\ .
\label{dop2.2}
\eea
Putting the result (\ref{dop2.2}) into the integrand in equation (\ref{tHnun1}) and
performing the integration over $w$ (again using the integral given 
in footnote (\ref{intdy})), we obtain
\bea
{\tH}_{\nu+1}(\sigma)^{\rm (MA,1-\ell)} = 
{\ntH}_{\nu+1}(\sigma)^{\rm (MA,1-\ell)} &=&
\frac{1}{(\delta+n) \pi \beta_0^{\delta+n+1}} 
\ {\rm Im} \frac{1}{\left[ \ln (\sigma/\bL^2) -i\pi \right]^{\delta+n}}
\label{dop2.3}
\\
&=& 
\frac{1}{\nu \pi \beta_0^{\nu+1}} 
\ \frac{1}{\left[ \ln^2 (\sigma/\bL^2) +\pi^2 \right]^{\nu/2}} \
\sin \left( \nu \; {\rm artan}(\pi/\ln (\sigma/\bL^2)) \right)
\ .
\label{dop2.4}
\eea
where $\nu=n+\delta$, with $0 < \delta < 1$ and $n=0,1,2,\ldots$.
The expression (\ref{dop2.4}) is explicit and is, by (analytic in $\nu$)
continuation valid for any $\nu$. It coincides with the result of BMS in 
\cite{BMS2}\footnote{
Their ${\tH}_{\nu+1}(\sigma)$ is our $\beta_0^{\nu+1} {\tH}_{\nu+1}(\sigma)$, where
$\beta_0=(1/4)(11 - 2 n_f/3)$.} 
and has, therefore, several interesting
properties derived and specified on \cite{BMS2} (their equations (3.12)-(3.13)
and (3.16)-(3.19)). Similarly as we did at the end of sec.~\ref{sec:man}
for the spacelike coupling, we can now use this explicit MA one-loop 
timelike coupling expression (valid now for any $\nu$) in order to 
extend the $\nu$-regime of applicability of the general anQCD 
time-like coupling formula (\ref{tHnu1}) [or: (\ref{tHnun1}) with $n=0$]
down to $\nu \approx -2$
\bea
{\tH}_{\nu+1}(\sigma) &=& 
{\tH}_{\nu+1}(\sigma)^{\rm (MA,1-\ell)} +
\frac{\sin (\pi \nu)}{\pi^2 \nu \beta_0^{\nu}}
\int_0^{\infty} \frac{d w}{w^{\nu}} 
\left[ \rho_1(\sigma e^w) - \rho_1(\sigma e^w)_{\rm pt}^{\rm (1-\ell)} \right]
\quad (-2 < \nu < 1) \ ,
\label{tHnu1b}
\eea
where ${\tH}_{\nu+1}(\sigma)^{\rm (MA,1-\ell)}$ is given in equation (\ref{dop2.4}), 
and $\rho_1(\sigma)_{\rm pt}^{\rm (1-\ell)}$ in equation (\ref{rho11l}). 
We can apply the limit $\nu \to -1$ in equation (\ref{tHnu1b}), and obtain
${\tH}_0(\sigma) = 1$ (for all $\sigma \geq 0$), because
${\tH}_{0}(\sigma)^{\rm (MA,1-\ell)} \equiv 1$.

\subsection{General form for the spacelike and the timelike observables}
\label{subsec:general}

We wish to stress that the formulas (\ref{disptAn3}), (\ref{disptAn3b}) and
(\ref{tHnu1}), (\ref{tHnun1}), (\ref{tHnu1b}) allow us to calculate the corresponding
couplings $\tA_{\nu+1}$ and $\tH_{\nu+1}$ for any real $\nu > -2$ and in any 
analytic QCD theory in which we know the discontinuity
function $\rho_1(\sigma) = {\rm Im} \A_1(Q^2=-\sigma - i \varepsilon)$
(or equivalently: the coupling function $\A_1(Q^2)$).

We can define the combinations of $\tA_{\nu +n}(Q^2)$'s\footnote{
We recall that the latter are defined via the integrals of 
equation (\ref{disptAn3}) or (\ref{disptAn3b}).} 
which are analogous to the pQCD relations 
(\ref{anutanu}) under the correspondence (\ref{anrule2})
\be
\A_{\nu} \equiv {\tA}_{\nu} + \sum_{m \geq 1}
\tk_m(\nu) {\tA}_{\nu + m} \quad (\nu > -2) \ .
\label{AnutAnu}
\ee
As expected, it is easy to check that the analytic series (\ref{Fmana})
can then be rewritten in the form
\be
{\cal F}(Q^2)_{\rm man} =
{\cal F}(Q^2)_{\rm an} \equiv {\A}_{\nu_0}(Q^2) + 
{\cal F}_1 {\A}_{\nu_0+1}(Q^2) + 
{\cal F}_2 {\A}_{\nu_0+2}(Q^2) + \cdots \ .
\label{Fana}
\ee
Therefore, the comparison with the original perturbation series in powers of
$a_{\rm pt}(Q^2)$, equation (\ref{calFpt}), gives us the correspondence between
the pQCD and anQCD quantities
\be
a_{\rm pt}^{\nu+1} \mapsto \A_{\nu+1}
\label{anrule3}
\ee
for any real, in general noninteger, $\nu > -2$.
Using the same combinations for the timelike couplings $\tH_{\nu+n}$'s
\be
\ntH_{\nu} \equiv {\tH}_{\nu} + \sum_{m \geq 1}
\tk_m(\nu) {\tH}_{\nu + m} \quad (\nu > -2) \ ,
\label{HnutHnu}
\ee
we can rewrite the associated timelike observable ${\cal T}(\sigma)$ of
equation (\ref{TF}) in a form similar to the expansion (\ref{calTman}) but
involving the original ${\cal F}_j$ coefficients instead of ${\widetilde {\cal F}}_j$
\be
{\cal T}(\sigma)_{\rm man} = {\cal T}(\sigma)_{\rm an} 
\equiv {\ntH}_{\nu_0}(\sigma) + 
{\cal F}_1 {\ntH}_{\nu_0+1}(\sigma) + 
{\cal F}_2 {\ntH}_{\nu_0+2}(\sigma) + \cdots \ .
\label{calTan}
\ee 
In equations (\ref{Fana}) and (\ref{calTan}), the subscript 'an' now
stands for ``analytic''.

Equation (\ref{AnutAnu}) further implies that $\A_0(Q^2) = 1$ (for all
complex $Q^2$), because: (a) ${\tA}_0(Q^2)=1$ as shown at the end of section
\ref{sec:lognonder}; and (b) ${\tk}_m(0)=1$ as seen from the results
(\ref{tkmsk1})-(\ref{tkmsk4}). Analogously, equation (\ref{HnutHnu})
implies ${\ntH}_0(\sigma)=1$ (for $\sigma \geq 0$), since ${\tH}_0(\sigma)=1$
as shown at the end of subsection \ref{subsec:timelike}. This means
that $\A_0 \equiv 1$ and ${\ntH}_0 \equiv 1$ in any anQCD, and this represents
a consistency check of our method 
of construction of $\A_{\nu}$ and $\ntH_{\nu}$.\footnote{We thank 
S.V.~Mikhailov for
pointing this out.}

The results of our method allow us to obtain also analytization of
powers combined with logarithms of the coupling
\bea
a_{\rm pt}^{\nu}(Q^2) \ln^k a_{\rm pt}(Q^2) & = & 
\frac{\partial^k a_{\rm pt}^{\nu}(Q^2)}{\partial \nu^k} \Rightarrow
\nonumber\\
\A_{\nu,k}(Q^2) \; \left[
\equiv \left( a^{\nu}(Q^2) \ln^k a(Q^2) \right)_{\rm an} \right]
& = & \frac{\partial^k \A_{\nu}(Q^2)}{\partial  \nu^k} \qquad
(\nu > -2; \; k=0,1,2,\ldots)
\label{AnulnkA}
\eea
The right-hand side  of equation (\ref{AnulnkA}) follows from the
analytization rule (\ref{anrule3}), applied separately to each
power in the expression $(a_{\rm pt}^{\nu+\delta} - a_{\rm pt}^{\nu})/\delta$
(where $\delta \to 0$) when $k=1$; when $k \geq 2$, the principle is the same.
The derivative $\partial^k/{\partial  \nu^k}$ on the right-hand side of
(\ref{AnulnkA}) is applied to each term on the right-hand side
of the sum (\ref{AnutAnu}), where we have to take into account that
the $\nu$-dependence is in the coefficients ${\tk}_m(\nu)$ and in the
couplings $\tA_{\nu+m}(Q^2)$ whose expression 
is given in equation (\ref{disptAn3b}).

\section{Application to the Higgs decay width}
\label{sec:appl}

Following references \cite{BMS1,BMS2,BMS3}, in
this section we apply the presented approach to the evaluation
of the decay width of the (Standard Model) Higgs into heavy
quark-antiquark ($b{\bar b}$) pair: $\Gamma(H \to b {\bar b})$
\be
\Gamma(H \to b {\bar b})(s) = \frac{N_c G_F}{4 \pi \sqrt{2}} \sqrt{s} \ T(s) \ ,
\label{GH1}
\ee
where $G_F$ is the Fermi coupling constant, $s=M_H^2$ is the square of
the Higgs mass, and $T(s)$ is the imaginary part 
${\rm Im} \Pi(-s-i \epsilon)/(6 \pi s)$
of the correlator of the scalar current $J_b=m_b {\bar b} b$
\be
\Pi(Q^2) = i (4 \pi)^2 \int dx \exp (i q x) 
\langle 0 | T [J_b(x) J_b(0)] | 0 \rangle \ ,
\label{Pi}
\ee
where $Q^2=-q^2$, cf.~\cite{Djouadi,Kataev}. 
Later on in this section, we will see that the perturbation expansion
of the corresponding spacelike quantity involves noninteger powers of $a_{\rm pt}$,
due to the $b$ quark mass anomalous dimension.
Using the notations of \cite{CKS}, 
we can write the timelike quantity $T(s)$ as a perturbation expansion
\be
T(s) = {\overline m}_b^2(s) \left( 1 + \sum_{j=1}^{\infty} t_n a_{\rm pt}^n(s) \right) \ ,
\label{Ts}
\ee
where the square of the (spacelike) renormalization scale $\mu$ 
was chosen to be $\mu^2=s$, and ${\overline m}_b(\mu^2)$ is the
${\overline {\rm MS}}$ running mass of the $b$ quark.
The corresponding spacelike quantity $F(Q^2)$ is
\be
F(Q^2) = Q^2 \int_0^{\infty} \frac{d \sigma \ T(\sigma)}{(\sigma + Q^2)^2}
\ , 
\label{FQ2}
\ee
and its expansion is written as
\be
F(Q^2) = {\overline m}_b^2(Q^2) \left(
1 + \sum_{j=1}^{\infty} f_n a_{\rm pt}^n(Q^2) \right)
\ ,
\label{FQ2exp}
\ee
Relations between the (dimensionless) coefficients $f_j$ and $t_j$ are given in \cite{CKS}. 

The idea is to evaluate first the spacelike quantity $F(Q^2)$, and obtain the timelike quantity $T(s)$ (and thus the decay width) by application
of the integral tranformation inverse to (\ref{FQ2}) 
[cf.~also equations (\ref{FT})-(\ref{TF})]
\be
T(\sigma) = \frac{1}{2 \pi i} 
\int_{-\sigma - i \varepsilon}^{-\sigma + i \varepsilon} 
\frac{d Q^{' 2}}{Q^{' 2}} F(Q^{' 2}) \ ,
\label{TF2}
\end{equation}

\subsection{Running mass}

For this, we will use, in the expression (\ref{FQ2exp}), for the square
of the running mass an expansion in (noninteger) powers of $a_{\rm pt}(Q^2)$. 
We recall that the renormalization group equation (RGE) for the 
squared ${\overline {\rm MS}}$ running mass is 
\be
\frac{d{\overline m}}{d \ln \mu^2} \equiv 
- {\overline m} \ \gamma_m(a_{\rm pt}) = 
- {\overline m} a_{\rm pt} 
\left( 1 + \sum_{j\geq 1} \gamma_j a_{\rm pt}^j \right) \ ,
\label{RGEm}
\ee
where the coefficients $\gamma_j$ ($j=1,2,3$) of the mass anomalous
dimension are known (\cite{massan2,massan3,massan4}); for
$n_f=5$, which applies in the case of the considered decay, we have:
$\gamma_1=3.51389$, $\gamma_2=7.41986$, $\gamma_3=11.0343$. The 5-loop
coefficient $\gamma_4$ has not yet been calculated. Nonetheless, application
of Pad\'e approximants to the quark mass anomalous dimension $\gamma_m(a_{\rm pt})$ 
for $n_f=5$ indicates that
$\gamma_4 \approx 12.$, and we will use this value.\footnote{
Namely, applying to $\gamma_m(a_{\rm pt})$ (at $n_f=5$) the Pad\'e 
approximants $[3/1](a)$, $[2/2](a)$, $[1/3](a)$, and reexpanding 
in powers of $a_{\rm pt}$ up to $a_{\rm pt}^5$, gives us
$\gamma_4=16.4, 9.2, 10.2,$ respectively; the arithmetic average is $11.9$,
i.e., approximately $12$. If repeating the same procedure at one order lower,
we obtain from Pad\'e approximants $[2/1]$ and $[1/2]$ the values
$\gamma_3=15.7, 8.8$, respectively, the average being $12.2$ which compares
favorably with the exact value $\gamma_3=11.0343$. The latter test
gives us reason to except that $\gamma_4=12.$ is a reasonable estimate.}
Furthermore, the $\beta_j$ ($j=0,1,2,3$) 
coefficients (in the ${\overline {\rm MS}}$
scheme) of the RGE (\ref{RGE}) for the coupling $a_{\rm pt}(Q^2)$ have been
calculated explicitly, \cite{beta0,beta1,beta2,beta3},
and for $n_f=5$ their values are: $\beta_0=1.91667$, $\beta_1=2.41667$,
$\beta_2=2.82668$, $\beta_3=18.8522$. The 5-loop beta coefficient has been
estimated in \cite{EJTKS} by Pad\'e-related methods, and for
$n_f=5$ the estimated value is $\beta_4=165.161$, which we will use here.

Integration of the RGE's (\ref{RGE}) and (\ref{RGEm}) gives for the
squared running mass the solution
\be
{\overline m}_b^2(\mu^2) = {\hat m}_b^2 \; a_{\rm pt}^{\nu_0}(\mu^2) 
\left( 1 + \sum_{j \geq 1} {\cal M}_j a_{\rm pt}^j(\mu^2) \right)
\label{barm2run}
\ee
where ${\hat m}_b^2$ is a renormalization scale invariant mass,
$\nu_0 = 2/\beta_0=1.04348$, and the coefficients ${\cal M}_j$ 
($j=1,2,3,4$) are functions of $\beta_0$, $c_k \equiv \beta_k/\beta_0$ and 
$\gamma_k$ ($k\leq j$), and they are given in appendix \ref{App4}. 
For $n_f=5$ these coefficients are: ${\cal M}_1=2.35098$;
${\cal M}_2=4.38319$; ${\cal M}_3=3.87308$; ${\cal M}_4=-22.2155$.
The mass 
quantity ${\hat m}_b$ is RG-invariant and can be obtained with high
precision in the following way. The world average value of
the QCD coupling parameter (in ${\overline {\rm MS}}$ scheme)
is $a_{\rm pt}(M_Z^2) \equiv 0.1184/\pi$ \cite{PDG2010}. The value of the
running mass at its own renormalization scale, 
${\overline m}_b( {\overline m}_b^2 )$ can be extracted from the 
heavy quarkonium physics. We will take the (central) value obtained in
\cite{CCG}: ${\overline m}_b( {\overline m}_b^2; n_f=4) = 4.24$ GeV. 
If taking the threshold between $n_f=4$ and $n_f=5$ at 
$\mu=4.24$ GeV, the aforementioned mass value and the
world average value $a(M_Z^2)= 0.1184/\pi$ \cite{PDG2010}
lead to the values\footnote{The discontinuity in the mass value
at threshold can be obtained from \cite{CKS98}.} at $n_f=5$
\be
{\overline m}_b( {\overline m}_b^2; n_f=5) = 4.232 \ {\rm GeV} \ ,
\quad
a_{\rm pt}({\overline m}_b^2; n_f=5) = 0.22542/\pi \ .
\label{mbarmb}
\ee
Using these values in the relation (\ref{barm2run}), we obtain the
scale invariant mass 
\be
{\hat m}_b = 15.330 \ {\rm GeV} \ .
\label{hatmb}
\ee
On the other hand, the values of the coefficients $f_j$ of the expansion
(\ref{FQ2exp}) for $j=1,2,3$ were obtained in \cite{f1f2f3}, and
for $j=4$ in \cite{f4} (denoted as ${\tilde d}_4$ there):
$f_1=5.66667$; $f_2=51.5668 - 1.90696 n_f$; $f_3=648.709 - 63.7418 n_f + 0.929133 n_f^2$;
$f_4=9470.76 - 1454.28 n_f + 54.7826 n_f^2 - 0.453744 n_f^3$. The values for the
here relevant case $n_f=5$ are: $f_1=5.66667$; $f_2=42.032$; $f_3=353.229$; 
$f_4=3512.2$; respectively. 

\subsection{Higgs decay}

We can now define the dimensionless (``reduced'') spacelike
quantity by dividing the expression $F(Q^2)$
[equations (\ref{FQ2}), (\ref{FQ2exp})]
by the RG-invariant scale ${\hat m}_b^2$, and using the expansion
(\ref{barm2run}) 
\bea
{\cal F}(Q^2) & \equiv & \frac{F(Q^2)}{{\hat m}_b^2} =
a(Q^2)^{\nu_0} + \sum_{n \geq 1} {\cal F}_n a^{\nu_0+n}(Q^2) \ ,
\label{calF}
\eea
where the coefficients ${\cal F}_n$ are now the corresponding
combinations of the coefficients $f_j$ and ${\cal M}_k$
\be
{\cal F}_n = f_n + f_{n-1} {\cal M}_1 + \cdots f_1 {\cal M}_{n-1} +
{\cal M}_n \ .
\label{calFn}
\ee
For $n_f=5$ this gives: ${\cal F}_1=8.01764$; ${\cal F}_2=59.7374$;
${\cal F}_3=480.756$; ${\cal F}_4=4526.6$.
The expression ${\cal F}(Q^2)$ of equation (\ref{calF}) is now the
expansion of a spacelike quantity in noninteger powers 
(with $\nu_0=2/\beta_0=1.04348$) considered in the
previous section, cf.~equations (\ref{calFpt}), (\ref{Fmpta}), (\ref{Fmana}).
The corresponding timelike quantity ${\cal T}(s)$ is
\be
{\cal T}(s) \equiv \frac{T(s)}{{\hat m}_b^2} =
\frac{\Gamma(H \to b {\bar b})(s)}{{\hat m}_b^2 N_c G_F \sqrt{s}/(4 \pi \sqrt{2})}
\label{calT}
\ee
where $s=M_H^2$ and we used the relation (\ref{GH1}). This quantity is
then evaluated by the formula (\ref{calTman}), with ${\widetilde {\cal F}}_n$'s
($n=1,2,3,4$)
determined by the relations (\ref{tFjFj}), as explained in the previous
section; or, equivalently, evaluated by the formula (\ref{calTan}). 
The evaluation can be performed in any analytic QCD theory,
and even in perturbative QCD, simply by using in expressions
(\ref{tHnu1}) and (\ref{tHnun1}) the discontinuity function of
the theory $\rho_1(\sigma) = {\rm Im} \A_1(-\sigma - i \epsilon)$ (in anQCD)
or $\rho_1(\sigma) = {\rm Im} a_{\rm pt}(-\sigma - i \epsilon)$ (in pQCD).

\subsection{Numerical calculations}

For numerical illustration of our approach, we will consider here 
the discontinuity function $\rho_1(\sigma)$ to originate: (a) from 
the Minimal Analytic (MA) model of Shirkov and Solovtsov 
\cite{ShS,MSS,Sh} 
(also known as Analytic Perturbation Theory - APT); 
(b) the models which have, at high $\sigma \geq M_H^2$, the same
$\rho_1(\sigma)$ as the perturbative QCD -- this includes analytic QCD
models of the type \cite{CCEM,Alekseev}, and the perturbative QCD itself.
We could contruct, in principle, such discontinuity functions
by numerically integrating the RGE for $a_{\rm pt}(Q^2)$ (in ${\overline {\rm MS}}$ 
renormalization scheme and with an initial condition at $Q^2=M_Z^2$) 
over the complex plane of $Q^2$ and evaluating the imaginary part
over the negative semiaxis. However, such an approach is cumbersome.
We calculate $\rho_1(\sigma)$ by evaluating $a_{\rm pt}(Q^2)$ for complex
$Q^2$ as a sum of the exact two-loop solutions $a_{\rm pt}(Q^2,2-\ell.)$
(which involve Lambert function, cf.~\cite{Gardi:1998qr,Magr}) 
as described in \cite{Kourashev}
\be
a_{\rm pt}(Q^2) = a_{\rm pt}(Q^2,2-\ell.)+
\sum_{j=3}^6 {\cal C}_j a_{\rm pt}^j(Q^2,2-\ell.) \ ,
\label{aQ2}
\ee
where 
\be
{\cal C}_3 = c_2, \quad {\cal C}_4 = \frac{1}{2} c_3, \quad
{\cal C}_5 = \left( \frac{5}{3} c_2^2 - \frac{1}{6} c_1 c_3 +
\frac{1}{3} c_4 \right),  \quad
{\cal C}_6 = \left[ \frac{1}{12} \left( - c_1 c_2^2 + c_1^2 c_3 
- 2 c_1 c_4 \right) + 2 c_2 c_3 + \frac{1}{4} c_5 \right] \ .
\label{calCs}
\ee
We truncate the series at $j=6$. The last coefficient ${\cal C}_6$ 
depends also on $c_5 = \beta_5/\beta_0$, which we do not know in
${\overline {\rm MS}}$ scheme (even the estimates are not reliable), so we
set $c_5=0$. Since the Lambert function can be called upon in various
numerical softwares, including Mathematica \cite{Math8}, this high
precision evaluation of $\rho_1(\sigma) = {\rm Im} a_{\rm pt}(-\sigma - i \epsilon)$ is fast.
The two-loop coupling in terms of the Lambert function 
$W_{\pm 1}$, \cite{Gardi:1998qr,Magr}, is 
\be
a_{\rm pt}(Q^2,2-\ell.) =
- \frac{1}{c_1} \frac{1}{\left[ 1 + W_{\mp 1} ( z ) \right]} \ ,
\label{apttH}
\ee
where $Q^2 = |Q^2| \exp( i \phi)$, the upper subscript refers to the case
$0 \leq \phi < + \pi$, the lower subscript to $- \pi < \phi < 0$, and
\be
z =  - \frac{1}{c_1 e} \left( \frac{|Q^2|}{\Lambda^2} \right)^{-\beta_0/c_1} \exp \left[ - i   \frac{\beta_0}{c_1} \phi  \right] \ .
\label{zexpr}
\ee
The Lambert scale $\Lambda$ at $n_f=5$ is $\Lambda=0.2642$ GeV in order for
the expansion (\ref{aQ2}) to reproduce the world average value 
$a_{\rm pt}(M_Z^2) = 0.1184/\pi$ [the corresponding usual
${\overline {\rm MS}}$ scale (at $n_f=5$) is
${\overline \Lambda} = 0.213$ GeV]. These values were used in our
evaluations. 

On the other hand, in the MA model \cite{ShS,MSS,Sh},
the value ${\overline \Lambda}_{\rm MA} = 0.260$ GeV (at $n_f=5$) is the one that
reproduces the high energy QCD phenomenology (see also \cite{BMS2}).
This value of ${\overline \Lambda}_{\rm MA}$ corresponds here to the 
Lambert scale value in MA
${\Lambda}_{\rm MA} = (0.260/0.213) {\Lambda} = 0.3225$ GeV. In MA, the
renormalization scale invariant mass ${\hat m}_b^2$ can be obtained
by replacing in equation (\ref{barm2run}) the noninteger powers
$a({\overline m}_b^2)^{\nu_0 + j}$ (where $j=0, \ldots, 4$) 
by $\A^{\rm (MA)}_{\nu_0 + j}({\overline m}_b^2)$,
the latter calculated via relations (\ref{AnutAnu}) and (\ref{disptAn3})
using for ${\overline m}_b$ the value of equation (\ref{mbarmb})
and for the discontinuity function the perturbative expression
$\rho_1^{\rm (pt)}(\sigma)$ with the Lambert scale value
$\Lambda_{\rm MA} = 0.3225$ GeV instead of $\Lambda=0.2642$ GeV.
This then results in the case of MA in a value of ${\hat m}_b$
somewhat lower than the one given in equation (\ref{hatmb})\footnote{
The value of ${\hat m}_b$ given in  equation (\ref{hatmb}) is the one
in pQCD and, to a large degree of precision, 
in all such analytic QCD models where the spacelike
analytic coupling merges fast with $a_{\rm pt}(Q^2)$ at high $Q^2$:
$\A_1(Q^2) - a_{\rm pt}(Q^2) \sim (\Lambda^2/Q^2)^n$ at $Q^2 \gg \Lambda^2$,
with $n \geq 3$. One such model was constructed in \cite{CCEM},
another in \cite{Alekseev}, both with $n=3$. We note that in MA $n=1$.}
\be
{\hat m}_b^{\rm (MA)} =15.029 \ {\rm GeV} \ .
\label{hatmbMA}
\ee

On the other hand, the usual perturbative QCD (pQCD) approach in 
evaluating the mentioned decay width is obtained by using
the pQCD expansion (\ref{Ts}) in powers of $a(s)$ ($\mu^2=s \equiv M_H^2$),
with the overall factor ${\overline m}_b^2(s)$ there given by 
equation (\ref{barm2run}), and the coefficients $t_n$ obtained from
coefficients $f_i$, $\gamma_j$ and ${\cal M}_j$ by using the integral relation
(\ref{FQ2}). On both sides of equation (\ref{FQ2}) expansions in powers of
$a(\mu^2)$ at the fixed renormalization scale $\mu^2=Q^2$ are used, 
and  ${\overline m}_b^2(s)/{\overline m}_b^2(Q^2)$ is also
expanded in powers of $a(Q^2)$. This then involves integrations of powers
of (large) logarithms $\ell=\ln (s/Q^2)$
\be
I_n \equiv Q^2 \int_0^{\infty}  \frac{d s \ \ln^n (s/Q^2)}{(s + Q^2)^2} \ ,
\label{lnints}
\ee
which are: $I_2=\pi^2/3$, $I_4= 7 \pi^4/15$, etc.; $I_{2 k +1}=0$).
For details, see \cite{CKS} and \cite{BMS3} (App.~A there), 
and their relations between $t_n$'s and $f_k$'s and
$\gamma_j$'s [their equations (22)-(24)]. 
At $n_f=5$, the coefficients $t_n$ and $f_n$ compare: 
$(f_n,t_n) = (5.66667, 5.66667)$; $(42.032, 29.1467)$; $(353.229, 41.7576)$;
$(3512.2, -825.747)$; for $n=1,2,3,4$, respectively. Here we see that the
effects of $I_{2 k}$ integrals tend to decrease the absolute values
of $|t_n|$ in comparison to $f_n$ for $n \leq 4$, and this makes
the pQCD evaluation (\ref{Ts}) numerically very well behaved.
However, there appears to exist no reason for this tendency 
to persist at higher orders. Further, looking at the integrals 
(\ref{lnints}), we see that they involve integration over 
large RGE logarithms.

\begin{figure}[htb] 
\centering\epsfig{file=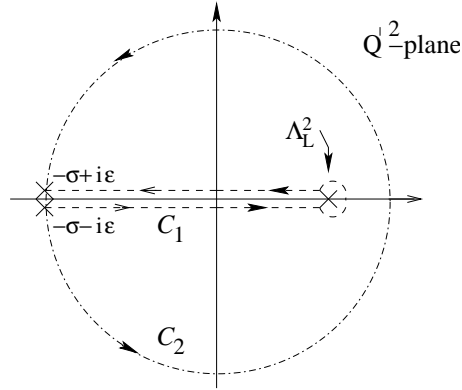,width=60mm}
\vspace{-0.4cm}
 \caption{\footnotesize  Paths ${\cal C}_1$ and ${\cal C}_2$
in the complex $Q^{' 2}$-plane for the case when the coupling has
unphysical (Landau) cut.}
\label{contour12p}
 \end{figure}

The described pQCD method, i.e., the 
power series for $T(\sigma)$ of equation (\ref{Ts}) 
truncated at $j=4$,
can be derived alternatively in the following way.
We reorganize the power series for $F(Q^{' 2})$ of
equation (\ref{FQ2exp}) into the form (\ref{calF})
in powers $a_{\rm pt}(Q^{' 2})^{\nu_0 + j}$ and include
the terms up to $j=4$. Then we put the resulting (truncated) series 
of $F(Q^{' 2})$ into the integral (\ref{TF2}) for $T(\sigma)$
where the integration path in the complex $Q^{' 2}$ plane is taken 
along the circular contour of radius $\sigma$ 
(the contour ${\cal C}_2$ in fig.~\ref{contour12p}).
However, instead of integrating the powers $a_{\rm pt}(Q^{' 2})^{\nu_0 + j}$
as they are [with $Q^{' 2} = \sigma \exp(i \phi)$ running], 
we use the perturbative RGE expansion of
these powers around the fixed scale $Q^{' 2} = \sigma > 0$. This sometimes
may not be the best approach, because the perturbative RGE expansion
involves powers of relatively large logarithms $\ln (Q^{' 2}/\sigma)
= i \phi$, with $- \pi < \phi < \pi$. We thus obtain a series
in powers $a_{\rm pt}(\sigma)^{\nu_0+j}$ which we truncate at $j=4$.
Reorganizing this series into the form with the overall
factor ${\overline m}_b^2(\sigma)$, we then obtain the series for $T(\sigma)$ 
of the form of equation (\ref{Ts}) truncated at $n=4$, where the 
coefficients $t_n$ turn out to be the aforementioned expressions
involving $f_k$'s and $\gamma_j$'s.    

\begin{figure}[htb] 
\begin{minipage}[b]{.49\linewidth}
\centering\epsfig{file=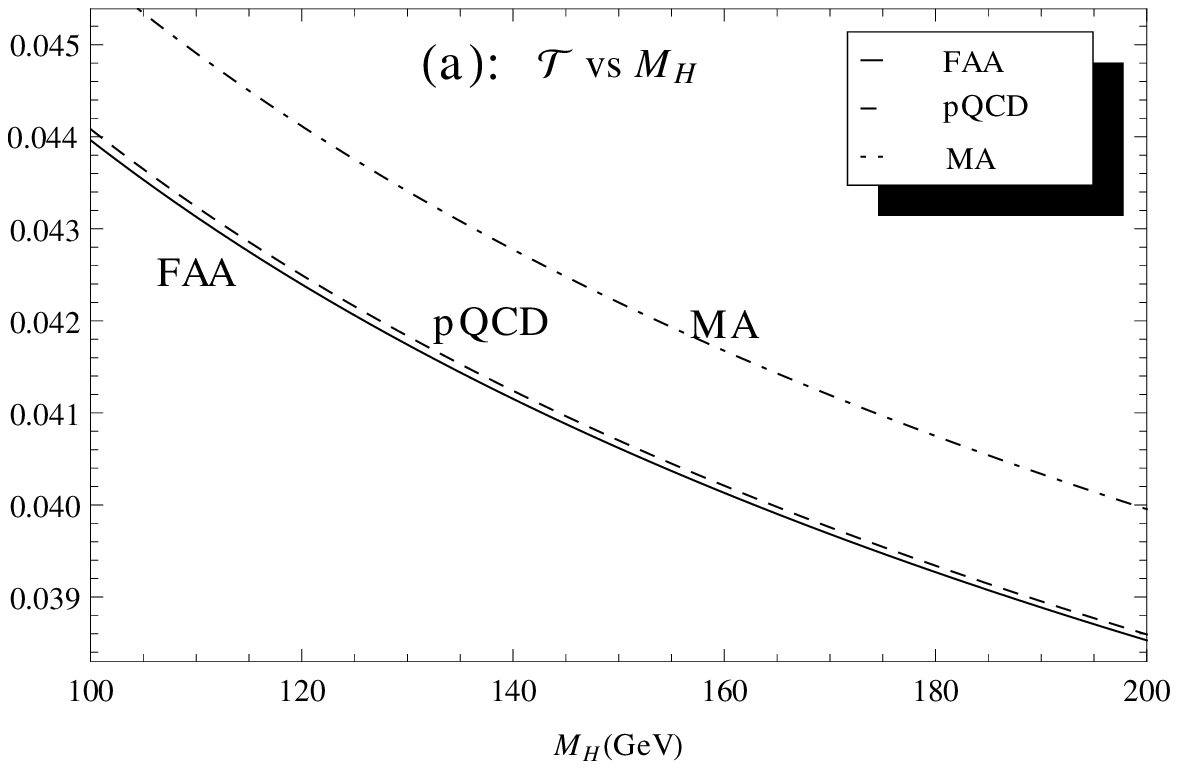,width=85mm}
\end{minipage}
\begin{minipage}[b]{.49\linewidth}
\centering\epsfig{file=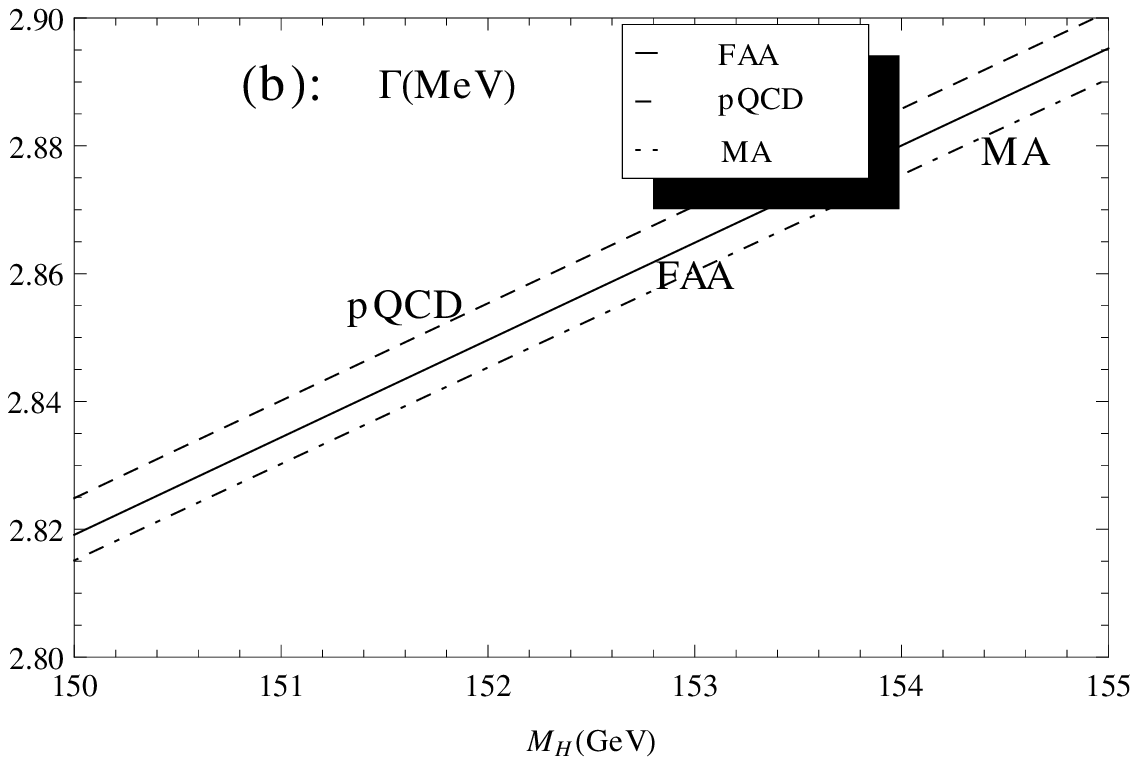,width=85mm}
\end{minipage}
\vspace{-0.4cm}
 \caption{\footnotesize  (a) 
The (dimensionless) quantity ${\cal T}(s)$, as defined in equation (\ref{calT}), 
as a function of the Higgs mass $M_H =\sqrt{s}$: 
for our approach (FAA) of equation (\ref{calTman}), for the
pQCD approach of equation (\ref{Ts}) 
(in both cases ${\overline \Lambda} = 0.213$ GeV at $n_f=5$); 
and for the minimal analytic (MA) model with FAA approach 
(${\overline \Lambda}_{\rm MA} = 0.260$ GeV at $n_f=5$);
(b) the same but now for the decay width $\Gamma(H \to b{\bar b})$.} 
\label{curv}
 \end{figure}

The just mentioned pQCD method of equation (\ref{Ts}) involves powers of
$a_{\rm pt}(\mu^2)$ at a fixed positive squared scale: 
$\mu^2 = \sigma = s = M_H^2$.
On the other hand, our approach,
equation (\ref{calTman}) or (\ref{calTan})[with ${\cal T}$ there defined via equation (\ref{calT})],
uses systematically the timelike quantities $\tH_{\nu+1}$ which are
contour-integrated couplings ${\widetilde \A}_{\nu+1}$ [cf.~equation (\ref{tHa})],
the latter corresponding to the
generalization of the logarithmic ``fractional'' (noninteger)
derivatives ${\widetilde \A}_{n+1}$
of the analytic coupling $\tA_1$ [equation (\ref{tAn})] 
in any analytic QCD model, or even in pQCD [equation (\ref{tan})].
We call our approch ``fractional analytic approach'' (FAA).
It can be applied to evaluation
of physical quantities [spacelike quantities ${\cal F}(Q^2)$, or
timelike quantities ${\cal T}(s)$] in any analytic QCD model. Further, it
can be applied formally even to evaluation of high energy timelike
quantities ${\cal T}(s)$ in pQCD, provided that $q^2=s$ is large enough:
$s > \Lambda^2_{\rm L}$ where $\Lambda^2_{\rm L}$ is the highest positive value
of the nonphysical (Landau) cut of $a_{\rm pt}(Q^2)$ 
in the complex $Q^2$-plane.
This is so because the contour integration in
fig.~\ref{contour12} in sec.~\ref{sec:man}
for equation (\ref{tHa}) can also be applied
to the pQCD coupling
\begin{equation}
{\tH}_{{\rm pt},\nu+1}(\sigma) \equiv \frac{1}{2 \pi i} 
\int_{-\sigma - i \varepsilon}^{-\sigma + i \varepsilon} 
\frac{d Q^{' 2}}{Q^{' 2}} {\widetilde a}_{{\rm pt}, \nu+1}(Q^{' 2}) \ ,
\label{tHpta}
\end{equation}  
provided the integration along the cut avoids the cut (including
its unphysical part), see the modified 
path ${\cal C}_1$ in fig.~\ref{contour12p} (in comparison to
the ${\cal C}_1$ in fig.~\ref{contour12}); and provided that at the same time
the integration along the circular path ${\cal C}_2$ gives the
same result -- the latter is the case only if $\sigma > \Lambda_L^2$.
The numerical results, for ${\cal T}(s)$ and $\Gamma(H \to b{\bar b})(s)$, 
as a function of the
Higgs mass $M_H = \sqrt{s}$, are presented in figs.~\ref{curv} a, b,
respectively.
We also include the curve for the Minimal Analytic (MA) model,
evaluated by our method and using ${\overline \Lambda}(n_f=5) =0.260$ GeV.
Our curve (FAA) can be interpreted as the result of application of
our method in perturbative QCD, or in any such analytic QCD in which the 
values of the discontinuity function
$\rho_1(\sigma) \equiv \A_1(-\sigma - i \epsilon)$ do not differ from the
values of the pQCD discontinuity function
$\rho_1^{\rm (pt)}(\sigma) \equiv a_{\rm pt}(-\sigma - i \epsilon)$ at
$\sqrt{\sigma} \geq M_H$.

We see that the FAA and pQCD curves are close to each other.
The MA curve for $\Gamma(H \to b{\bar b})$ comes close to the FAA and pQCD
curves because the effects of different values of scales ${\overline \Lambda}$
($0.260$ GeV instead of $0.213$ GeV) and different values of
${\hat m}_b^2$ [equations (\ref{hatmbMA}) instead of equation (\ref{mbarmb})]
tend to cancel each other.
The evaluation for the curves was performed by using the
renormalization scales such that $|\mu^2| = s$ ($=M_H^2$). If we vary the value
of the renormalization scale $|\mu^2|$ from $s/2$ to $2 s$ around
$s$, the values remain very stable; for example, when $M_H=150$ GeV,
the values of ${\cal T}$ are $(4.062 \pm 0.003) \cdot 10^{-2}$
in the FAA method case, and $(4.070 \pm 0.001) \cdot 10^{-2}$
in the pQCD method case.

The decay width, by FAA method, is $\Gamma(H \to b{\bar b}) = 2.03, 2.82, 3.57$ MeV,
when $M_H=100, 150, 200$ GeV, respectively.

It turns out that application of the expansion (\ref{calTan}) in our FAA
approach, instead of the expansion (\ref{calTman}), 
gives the same result.
Furthermore, the spacelike couplings $\A_{\nu}(Q^2)$ 
obtained via equations (\ref{AnutAnu}) and (\ref{disptAn3}), 
and the timelike couplings $\ntH_{\nu}(Q^2)$ obtained via
equations (\ref{tHnu1}) and (\ref{tHn1}) and (\ref{HnutHnu}),
when evaluated in such analytic QCD models
which have $\rho_1(\sigma) = {\rm Im} a_{\rm pt}(-\sigma - i \epsilon)$, 
i.e., in MA-type models,
give results numerically indistinguishable from the expressions
[see also equation (\ref{MAAndisp})]
\bea
\A_{\nu}^{\rm (APT)}(Q^2) &\equiv& \frac{1}{\pi} \int_0^{\infty} \frac{d \sigma \ 
{\rm Im} a_{\rm pt}^{\nu}(-\sigma - i \epsilon)}{\sigma + Q^2} \ ,
\label{AnuAPT}
\\
\ntH_{\nu}^{\rm (APT)}(\sigma) & \equiv & 
\frac{1}{\pi} \int_{\sigma}^{\infty} \frac{d \sigma^{\prime}}{\sigma^{\prime}}
{\rm Im} a_{\rm pt}^{\nu}(-\sigma^{\prime} - i \epsilon) \ ,
\label{HnuAPT}
\eea
for any positive $\nu$ (in general noninteger); and
$Q^2$ in the complex plane outside the negative semiaxis; and
$\sigma > 0$.
This is another check of consistency of our method, because it shows that the APT
construction, \cite{MSS,Sh}, of higher power analogs in MA 
of Shirkov and Solovtsov, when generalized from integer to noninteger
powers ($n \mapsto \nu$) as in equations (\ref{AnuAPT})-(\ref{HnuAPT}), gives the same
result as our method. 

Nonetheless, in conclusion, we stress that our method of
construction of spacelike couplings $\tA_{\nu}$ and $\A_{\nu}$
[equations (\ref{disptAn3}) and (\ref{AnutAnu})],
and of the corresponding timelike couplings $\tH_{\nu}$ and $\ntH_{\nu}$
[equations (\ref{tHnu1}), (\ref{tHn1}), (\ref{HnutHnu})] can be applied
also to any other analytic QCD models, e.g., models where
$\rho_1(\sigma) \not= {\rm Im} a_{\rm pt}(-\sigma - i \epsilon)$. The latter
inequality can be expected in general at low positive $\sigma$ values,
cf.~\cite{CCEM}. In such models, the APT-type construction,
equations (\ref{AnuAPT})-(\ref{HnuAPT}), or modifications thereof, cannot be applied.

We recall that the timelike couplings $\tH_{\nu}(\sigma)$ and ${\ntH}_{\nu}(\sigma)$
of our (FAA) method
depend only on $\rho_1(\sigma')$ at $\sigma' = \sigma e^w \geq \sigma$ --
see equations (\ref{tHnu1}), (\ref{tHn1}), (\ref{HnutHnu}). 
In the presented application of our method to $\Gamma(H \to b{\bar b})$,
however, $\sigma = s = M_H^2 > 100^2 \ {\rm GeV}^2$, i.e., only those 
$\rho_1(\sigma')$ contribute for which $\sigma'$ is very high ($> M_H^2$).
At such high $\sigma'$ we can expect that 
$\rho_1(\sigma') = {\rm Im} a_{\rm pt}(-\sigma^{\prime} - i \epsilon)$. Therefore,
in such cases the formulas (\ref{HnuAPT}) for ${\ntH}_{\nu}$'s can be applied 
[but not the formulas (\ref{AnuAPT}) for $\A_{\nu}$'s] and they
give the same result for $\ntH_{\nu}(\sigma)$ as our approach. 
Really, the FAA curve and the MA curve<
in figs.~\ref{curv} are numerically
reproduced by application of equations (\ref{HnuAPT}), using the
corresponding values ${\overline \Lambda}=0.213$ GeV and $0.260$ GeV, respectively. 

It would also be interesting to apply our method to evaluation of low-energy  
timelike observables, where $\rho_1(\sigma)$ may differ significantly from
the perturbative value. The method can also be applied to evaluation of
spacelike quantities.

\section{Conclusions}
\label{sec:concl}

We presented a method of calculating spacelike and timelike
QCD observables whose 
perturbation expansion in perturbative QCD (pQCD)
has noninteger powers of
the perturbative coupling $a_{\rm pt}$ $(\equiv \alpha_s/\pi)$.

The method can be applied in any analytic QCD model, i.e., 
(a) in any model with a given analytic spacelike coupling $\A_1(Q^2)$ 
(where $\A_1(Q^2)$ is the analytic analog of spacelike\footnote{
spacelike in the sense that $Q^2 (\equiv -q^2) \in \mathbb{C} \backslash (-\infty, 0]$}
$a_{\rm pt}(Q^2)$);
(b) or with a given discontinuity function 
$\rho_1(\sigma) \equiv {\rm Im} \A_1(-\sigma - i \epsilon)$ (where $\sigma \geq 0$).
Specifically, first we constructed the analytic analogs $\tA_{\nu+1}(Q^2)$
of the ($\nu$-noninteger extension) of the logarithmic derivatives
$\ta_{{\rm pt},\nu+1}(Q^2) \propto d^{\nu} a_{\rm pt}(Q^2)/d (\ln Q^2)^{\nu+1}$,
where $\nu$ can be any real number larger than $-2$, 
cf.~equations (\ref{disptAn3}), (\ref{disptAn3b}).
Furthermore, we constructed the corresponding timelike (Minkowskian) 
couplings $\tH_{\nu+1}(\sigma)$ ($\sigma \geq 0$), 
cf.~equations (\ref{tHnun1}), (\ref{tHnu1b}). Subsequently, we
obtained the analytic spacelike 
couplings $\A_{\nu}(Q^2)$ as a linear combination
of the aforementioned $\tA_{\nu+m}(Q^2)$'s ($m = 0, 1, 2, \ldots$), where
the couplings $\A_{\nu}(Q^2)$ are analytic analogs (in any given analytic
QCD models) of the powers $a_{\rm pt}(Q^2)^{\nu}$ ($\nu$ any real number
above $-1$), cf.~equation (\ref{AnutAnu}). 
Furthermore, the corresponding timelike (Minkowskian)
power analogs $\ntH_{\nu}(\sigma)$ were constructed, as the corresponding
linear combination of $\tH_{\nu+m}(Q^2)$ ($m = 0, 1, 2, \ldots$), 
cf.~(\ref{HnutHnu}).

We further demonstrated that in the Minimal Analytic model (MA, also named APT)
of Shirkov, Solovtsov, Solovtsova and Milton \cite{ShS,MSS,Sh},
our method gives the same explicit results for
$\A_{\nu+1}(Q^2)$ and $\tH_{\nu+1}(\sigma)$ at the one-loop level 
as the method of \cite{BMS1,BMS2,BMS3} of Bakulev, Mikhailov and Stefanis
(BMS; whose method can be applied in these MA-type models only), 
cf.~equations (\ref{disptAn31l}) and (\ref{dop2.3}). 
When going beyond the one-loop level within MA, the explicit 
formulas for $\A_{\nu+1}(Q^2)$ and $\tH_{\nu+1}(\sigma)$ in 
\cite{BMS1,BMS2,BMS3} become complicated and the comparison
with our results becomes harder. Numerically, though, we have
strong indications that within MA both BMS and our method agree also beyond
the one-loop level. We recall that our results are given in
form of integral, i.e., they are less explicit than the results of
\cite{BMS1,BMS2,BMS3} for MA. However, our results are 
applicable in any analytic QCD, and look simple in its (integral) form.
  
When the analytic QCD model is based on a beta function 
$\beta(\A_1)$ which is analytic at $\A_1=0$ (\cite{CKV}), i.e., 
in perturbative analytic QCD, we simply obtain $\A_n = \A_1^n$
\cite{CKV} and $\A_{\nu} = \A_1^{\nu}$.

Furthermore, our method can be applied to evaluation of timelike observables 
${\cal T}(s)$ within 
the nonanalytic pQCD, provided that $s > \Lambda^2_{\rm L}$, where
$\Lambda^2_{\rm L}$ is the positive branching point of the unphysical 
(Landau) $Q^2$-cut  $(0,\Lambda^2_{\rm L})$ of $a_{\rm pt}(Q^2)$.
This latter approach can be described as a
RGE-resummed contour method, as opposed to the more usual fixed-scale
contour method in pQCD. On the other hand, for spacelike observables,
while our method can be applied in (any) analytic QCD, 
it cannot be applied in nonanalytic pQCD, and the results are different 
from the usual (nonanalytic) pQCD evaluation results. 

Further, if we work in an analytic QCD model for which the 
discontinuity function $\rho_1(\sigma)$ is equal to its pQCD counterpart 
$\rho_1^{\rm (pt)}(\sigma)$ for large enough
$\sigma > M_0^2$ (where $M_0 \sim 1 \ {\rm GeV} > \Lambda_{\rm L}$, $M_0$ being
a typical scale of the onset of pQCD), 
then the method gives for timelike observables ${\cal T}(s)$ at 
$s > M_0^2$ the same result as the method 
gives in nonanalytic pQCD. For spacelike observables 
no analogous statement holds. This is so
because the spacelike couplings [$\A_1(Q^2)$, $\tA_{\nu}(Q^2)$, $\A_{\nu}(Q^2)$] are
represented by integrals involving the values of $\rho_1(\sigma)$ along
the entire cut of $\A_1(Q^2)$, while the timelike couplings [$\ntH_1(s)$,
$\tH_{\nu}(s)$, $\ntH_{\nu}(s)$] involve only $\rho_1(\sigma)$ for the
cut sector $\sigma \in (s,+\infty)$. 

We applied the method to evaluation of the Higgs decay width into
$b {\bar b}$ pair $\Gamma(H \to b{\bar b})$, as a function of the Higgs mass $M_H$. 
The results of this evaluation turn out to be the
same in pQCD and in any analytic QCD with $\rho_1(\sigma) = \rho_1^{\rm (pt)}(\sigma)$
at $\sigma \geq M_H^2$, because this is a high-energy timelike observable:
$\Gamma(H \to b{\bar b}) \propto {\cal T}(s)$ with $s=M_H^2 \gg \Lambda_{\rm L}^2$.

It would be also interesting to apply our method in analytic QCD models
to evaluation of low-energy timelike observables ${\cal T}(s)$ that 
involve noninteger powers, where $\rho_1(\sigma)$ at $\sigma \sim s$ 
may differ significantly from the perturbative value; 
and to evaluation of (low-energy) spacelike quantities in such models.
For the latter, we plan to investigate the structure functions of
the deep inelastic lepton-hadron scattering in analytic QCD models.

\begin{acknowledgments}
\noindent
G.C. is grateful to A.~P.~Bakulev, S.~V.~Mikhailov and D.~V.~Shirkov
for constructive comments.
This work was supported in part by FONDECYT (Chile) Grant No.~1095196 
(G.C. and A.K.), Rings Project (Chile) ACT119 (G.C.), 
and RFBR (Russia) Grant No.~10-02-01259-a (A.K.).
\end{acknowledgments}

\appendix

\section{Coefficients $k_m(\nu)$ and $\tk_m(\nu)$}
\label{App1}

\subsection{Coefficients $k_m(\nu)$: results}
\label{App1:km}

We need to find the coefficients $k_m(\nu)$, where $m=1,2,\ldots$,
and $\nu$ is any real number. These coefficients
are derived later in this appendix. First we write down their solution
explicitly.

It turns out that they involve derivatives 
$Z_m(\nu+n)$ of the Riemannian $\Gamma$-functions
\be
Z_m(\nu) \equiv \frac{1}{\Gamma(\nu+1)} \frac{d^m}{d x^m} 
{\left(\frac{\Gamma(\nu+1+x)}{\Gamma(1+x)}\right)}{\Big |}_{x=0} \ .
\label{Zm}
\ee
These functions can be expressed in the form of the Euler
$\Psi$-functions and their derivatives
\bea
\Psi(\nu) &\equiv& \frac{d}{d\nu} \ln \Gamma(\nu) \ , \quad
\Psi^{(m)}(\nu) \equiv  \frac{d^m}{d\nu^m} \Psi(\nu) \ .
\label{Psis}
\\
S_1(\nu) & \equiv & \Psi(\nu+1)-\Psi(1) \ , 
\label{S1}
\\
S_m(\nu) &\equiv& \frac{(-1)^{m-1}}{(m-1)!} \, 
\left(\Psi^{(m-1)}(\nu+1)-\Psi^{(m-1)}(1) \right) \quad
(m=2,3,\ldots) \ .
\label{Sms}
\eea
We note that $S_m(\nu)$ for integer $\nu=n$ (and $m=1,2,\ldots$) coincide
with the usual harmonic numbers (of order $m$):\footnote{
In Mathematica \cite{Math8}, $\Psi^{(m)}(\nu)$ is denoted
${\rm PolyGamma}[m,\nu]$; and $S_m(\nu)$ is denoted ${\rm HarmonicNumber}[n,m]$.}
$S_m(n) = \sum_{k=1}^n k^{-m}$.
In terms of these functions, the functions $Z_m(\nu)$ of equation (\ref{Zm})
are
\bea
Z_0(\nu) &=& 1 \ , \quad Z_1(\nu) = S_1(\nu) \ ,
\quad Z_2(\nu)  = S_1(\nu)^2 - S_2(\nu) \ ,
\nonumber \\
Z_3(\nu) &=& S_1(\nu)^3 -  3 S_1(\nu) S_2(\nu)
+2 S_3(\nu) \ ,  
\nonumber \\
Z_4(\nu) &=& S_1(\nu)^4 - 6 S_1(\nu)^2 S_2(\nu) + 3 S_2(\nu)^2
+ 8 S_1(\nu) S_3(\nu) - 6 S_4(\nu) \ ,
\label{Zms}
\eea
In terms of the quantities $Z_m(\nu)$, as given by equation (\ref{Zm}),
the coefficients $k_m(\nu)$ are
\bea
k_1(\nu) &=& \nu \, c_1 \, B_1(\nu) , 
\label{kmsk1}
\\
k_2(\nu) &=& \nu(\nu +1) \, 
\bigg(c_2 \, B_2(\nu) +  \frac{c_1^2}{2} \, B_{1,1}(\nu) \biggr), 
\label{kmsk2}
\\
k_3(\nu) &=& \frac{\nu(\nu +1)(\nu +2)}{2} 
\bigg(c_3 \, B_3(\nu) + c_1c_2 \, B_{1,2}(\nu) + 
 \frac{c_1^3}{3} \, B_{1,1,1}(\nu) \biggr), 
\label{kmsk3}
\\
k_4(\nu) &=& \frac{\nu(\nu +1)(\nu +2)(\nu +3)}{6} 
\bigg(c_4 \, B_4(\nu) + c_2^2 \, B_{2,2}(\nu) + 
\frac{c_1 c_3}{2} \, B_{1,3}(\nu) +
\frac{c_1^2 c_2}{2} \,  B_{1,1,2}(\nu) +\frac{c_1^4}{4} \, B_{1,1,1,1}(\nu) \biggr),
\label{kmsk4}
\eea
where  
\bea
B_1(\nu) &=& Z_1(\nu) - 1  \ , \quad
B_2(\nu) = \frac{\nu -1}{2(\nu +1)} \ , \quad
B_{1,1}(\nu) =  Z_2(\nu +1) - 2  \, Z_1(\nu) + 1 \ , 
\nonumber\\
B_3(\nu) &=& \frac{1}{6} - \frac{1}{\nu+1} + \frac{1}{\nu+2} \ , \quad
B_{1,2}(\nu) =  \frac{\nu}{\nu+2} \, Z_1(\nu +1)
+ \frac{2}{\nu +1} + \frac{1}{\nu +2} - \frac{11}{6} \ , 
\nonumber \\
B_{1,1,1}(\nu) &=& Z_3(\nu +2)-3 \, Z_2(\nu +1) + 
3  \, Z_1(\nu) - 1 \ , 
\nonumber\\
B_4(\nu) &=& \frac{1}{12} - \frac{2}{(\nu +1)(\nu +2)(\nu +3)} \ ,
\quad
B_{2,2}(\nu)  = \frac{5}{12} + \frac{1}{\nu+1} + \frac{1}{\nu+2} - 
\frac{5}{\nu+3} \ ,  
\nonumber \\
B_{1,3}(\nu)  &=& \left(1- \frac{6}{(\nu+2)(\nu+3)}\right) \, Z_1(\nu +1)
+ \frac{4}{\nu +1} - \frac{1}{\nu +2}   - \frac{1}{\nu +3} - \frac{13}{6}
 \ ,  
\nonumber \\
B_{1,1,2}(\nu) &=& \frac{3(\nu +1)}{\nu +3} Z_2(\nu +2) + 
\left(\frac{12}{\nu +2} + \frac{6}{\nu +3} - 11\right) \, Z_1(\nu +1)
- \frac{6}{\nu +1} - \frac{5}{\nu +2}   - \frac{11}{\nu +3} + \frac{38}{3}
 \ , 
\nonumber \\
B_{1,1,1,1}(\nu) &=&  Z_4(\nu +3) - 4 \, Z_3(\nu +2) + 6 \,
Z_2(\nu +1) - 4  \, Z_1(\nu) + 1 \ .
\label{Baux}
\eea

\subsection{Coefficients $\tk_m(\nu)$: results}
\label{App1:tkm}

Inversion of the relations (\ref{tanuanu}) gives us the expansion
(\ref{anutanu}) and the $\tk_m(\nu)$ coefficients in terms of the
previously written coefficients $k_{\ell}(\nu+n)$ ($\ell = 1,\ldots,m$,
$n=0,1,\ldots,m-1$)
\bea
\tk_1(\nu) & = & k_1(\nu) \ , \quad
\tk_2(\nu) =  k_1(\nu) k_1(\nu+1) - k_2(\nu) \ ,
\nonumber\\
\tk_3(\nu) &=& - k_1(\nu) k_1(\nu+1) k_1(\nu+2) + k_1(\nu) k_2(\nu+1)
+ k_1(\nu+2) k_2(\nu) - k_3(\nu) \ ,
\nonumber\\
\tk_4(\nu) &=&  k_1(\nu) k_1(\nu+1) k_1(\nu+2) k_1(\nu+3)
- k_1(\nu) k_1(\nu+1) k_2(\nu+2) - k_1(\nu) k_1(\nu+3) k_2(\nu+1)
\nonumber\\
&&
- k_1(\nu+2) k_1(\nu+3) k_2(\nu) 
+ k_2(\nu) k_2(\nu+2) + k_1(\nu) k_3(\nu+1) + k_1(\nu+3) k_3(\nu) - k_4(\nu) 
\ .
\label{tkms}
\eea
It is possible to check that our formulas, equations (\ref{kmsk1})-(\ref{kmsk4})
and (\ref{tkms}), give
\be
k_n(1) = \tk_n(1) = 0 \ ,
\label{nueq0}
\ee
reflecting the fact that, by definition [equations (\ref{tan}) and (\ref{tAn})],
$\tA_1 \equiv \A_1$ and $\ta_{{\rm pt},1} \equiv a_{\rm pt}$. 

It is interesting that the coefficients $\tk_m(\nu)$ 
can be cast in an equivalent alternative form which is somewhat similar 
to the expressions for the coefficients $k_m(\nu)$, 
equations (\ref{kmsk1})-(\ref{Baux}), 
and where instead of the derivatives (\ref{Zm}), 
another type of derivatives appears naturally: 
\be
\tZ_m(\nu) \equiv \Gamma(\nu+1) \frac{d^m}{d x^m} 
\left( \frac{\Gamma(1-x)}{\Gamma(\nu+1-x)} \right){\Big |}_{x=0} \ .
\label{tZm}
\ee
In terms of the harmonic numbers of order $k$, $S_k(n) = \sum_{s=1}^n s^{-k}$,
functions $Z_m(n)$ and $\tZ_m(n)$ can be expressed as
\bea
Z_m(n) &\equiv&  \frac{1}{\Gamma(n+1)} \frac{d^m}{d x^m} 
{\left(\frac{\Gamma(n+1+x)}{\Gamma(1+x)}\right)}{\Big |}_{x=0}
= \frac{d^m}{d x^m} \exp \left[ \sum_{k=1}^{\infty}
\frac{(-1)^{k+1} x^k}{k} S_k(n) \right] {\Big |}_{x=0} \ ,
\label{Zmexp}
\\
\tZ_m(n) &\equiv&  \Gamma(n+1) \frac{d^m}{d x^m} 
{\left(\frac{\Gamma(1-x)}{\Gamma(n+1-x)}\right)}{\Big |}_{x=0}
= \frac{d^m}{d x^m} \exp \left[ \sum_{k=1}^{\infty}
\frac{x^k}{k} S_k(n) \right] {\Big |}_{x=0} \ .
\label{tZexp}
\eea
This means that for $\tZ_m$'s, formulas very similar to those of
equations (\ref{Zms}) for $Z_m$'s, are valid
\bea
\tZ_0(\nu) &=& 1 (= Z_0(\nu)) \ , \quad 
\tZ_1(\nu) = S_1(\nu) (= Z_1(\nu)) \ ,
\quad \tZ_2(\nu)  = S_1(\nu)^2 + S_2(\nu) \ ,
\nonumber \\
\tZ_3(\nu) &=& S_1(\nu)^3 +  3 S_1(\nu) S_2(\nu)
+2 S_3(\nu) \ ,  
\nonumber \\
\tZ_4(\nu) &=& S_1(\nu)^4 + 6 S_1(\nu)^2 S_2(\nu) + 3 S_2(\nu)^2
+ 8 S_1(\nu) S_3(\nu) + 6 S_4(\nu) \ ,
\label{tZms}
\eea
The coefficients $\tk_m(\nu)$ are then written in a form very similar to
the formulas (\ref{kmsk1})-(\ref{kmsk4}) for $k_m(\nu)$'s
\bea
\tk_1(\nu) &=& - \nu \, c_1 \, \tB_1(\nu) , 
\label{tkmsk1}
\\
\tk_2(\nu) &=& \nu(\nu +1) \, 
\bigg(- c_2 \, \tB_2(\nu) +  \frac{c_1^2}{2} \, \tB_{1,1}(\nu) \biggr), 
\label{tkmsk2}
\\
\tk_3(\nu) &=& \frac{\nu(\nu +1)(\nu +2)}{2} 
\bigg(- c_3 \, \tB_3(\nu) + c_1c_2 \, \tB_{1,2}(\nu) - 
 \frac{c_1^3}{3} \, \tB_{1,1,1}(\nu) \biggr), 
\label{tkmsk3}
\\
\tk_4(\nu) &=& \frac{\nu(\nu +1)(\nu +2)(\nu +3)}{6} 
\bigg(- c_4 \, \tB_4(\nu) \! + \! c_2^2 \, \tB_{2,2}(\nu) \! + \! 
\frac{c_1 c_3}{2} \, \tB_{1,3}(\nu) \! - \!
\frac{c_1^2 c_2}{2} \,  \tB_{1,1,2}(\nu) \! + \!
\frac{c_1^4}{4} \, \tB_{1,1,1,1}(\nu) \biggr),
\label{tkmsk4}
\eea
where  
\bea
\tB_1(\nu) &=& \tZ_1(\nu) - 1  \ , \quad
\tB_2(\nu) = \frac{\nu -1}{2(\nu +1)} \ , \quad
\tB_{1,1}(\nu) =  \tZ_2(\nu) - 2  \, \tZ_1(\nu+1) + 1 \ , 
\nonumber\\
\tB_3(\nu) &=& 
\frac{1}{6} - \frac{1}{\nu+1} + \frac{1}{\nu+2} \ ,
\quad
\tB_{1,2}(\nu) =  \frac{\nu-1}{6(\nu+1)} 
\left(6 \tZ_1(\nu +1) -1 + \frac{4}{\nu +2}\right) \ , 
\nonumber \\
\tB_{1,1,1}(\nu) &=& \tZ_3(\nu)-3 \, \tZ_2(\nu +1) + 
3  \, \tZ_1(\nu+2) - 1 \ , 
\nonumber\\
\tB_4(\nu) &=& \frac{1}{12} - \frac{2}{(\nu +1)(\nu +2)(\nu +3)} \ ,
\quad
\tB_{2,2}(\nu)  = \frac{13}{12} - \frac{1}{\nu +1}-\frac{1}{\nu +2}
-\frac{1}{\nu+3}  \ ,
\nonumber \\
\tB_{1,3}(\nu)  &=& \left(1- \frac{6}{(\nu+1)(\nu+2)}\right) \, \tZ_1(\nu +3)
+ \frac{1}{6}+\frac{4}{\nu +1} - \frac{5}{\nu +2}   - 
\frac{2}{\nu +3} \ ,
\nonumber \\
\tB_{1,1,2}(\nu) &=& \frac{3(\nu -1)}{\nu +1} \tZ_2(\nu+2) - 
\left(1- \frac{6}{(\nu +1)(\nu +2)}\right) \, \tZ_1(\nu+3)
+ \frac{8}{3} -\frac{2}{\nu +1} + \frac{1}{\nu +2}   - \frac{8}{\nu +3} \ ,
\nonumber \\
\tB_{1,1,1,1}(\nu) &=&  \tZ_4(\nu) - 4 \, \tZ_3(\nu +1) + 6 \,
\tZ_2(\nu +2) - 4  \, \tZ_1(\nu+3) + 1 \ .
\label{tBaux}
\eea

\subsection{Reorganization of the power series; RS-dependence of new coefficients}
\label{App1:tildeFj}

The relation (\ref{anutanu}) between  $a_{\rm pt}^{\nu}$
and $\ta_{{\rm pt},\nu+m}$'s allows us to obtain immediately the
relation between the coefficients ${\cal F}_j$ of the
usual perturbative expansion (\ref{calFpt}) and the coefficients 
${\widetilde {\cal F}}_k$ of the reorganized (``modified'')
expansion (\ref{Fmpta})
\bea
{\widetilde {\cal F}}_1 & = & {\cal F}_1 + \tk_1(\nu_0) \ ,
\label{tF1}
\\
{\widetilde {\cal F}}_2 & = & {\cal F}_2 + \tk_1(\nu_0+1){\cal F}_1 +  \tk_2(\nu_0)
\ ,
\label{tF2}
\\
{\widetilde {\cal F}}_3 & = & {\cal F}_3 + \tk_1(\nu_0+2){\cal F}_2 +  \tk_2(\nu_0+1){\cal F}_1 +
\tk_3(\nu_0) \ ,
\label{tF3} \\
{\widetilde {\cal F}}_4 & = & {\cal F}_4 + \tk_1(\nu_0+3){\cal F}_3 +  \tk_2(\nu_0+2){\cal F}_2 +
\tk_3(\nu_0+1){\cal F}_1 + \tk_4(\nu_0)\ ,
\label{tF4}
\eea
All these are coefficients when the renormalization scale (RScl) $\mu^2$ is taken to
be $\mu^2=Q^2$. If using another RScl $\mu^2 \not= Q^2$ to calculate the observable 
${\cal F}(Q^2)$, we can obtain the RScl dependence of the coefficients
${\widetilde {\cal F}}_j(\mu^2/Q^2)$ by using the relations 
$d \tA_{\nu}(\mu^2)/d \ln \mu^2 = - \beta_0 \nu \tA_{\nu+1}(\mu^2)$ [or equivalently:
$d \ta_{{\rm pt},\nu}(\mu^2)/d \ln \mu^2 = - \beta_0 \nu \ta_{{\rm pt},\nu+1}(\mu^2)$],
cf.~equation \ref{tArecur}), and 
the RScl independence of the (spacelike observable) ${\cal F}(Q^2)$.
The resulting expressions are
\be
{\widetilde {\cal F}}_n(\mu^2/Q^2) = {\widetilde {\cal F}}_n +
\sum_{k=1}^n \frac{\Gamma(\nu_0+n)}{\Gamma(\nu_0+n-k)}
\frac{1}{k!} \beta_0^k \ln^k \left( \frac{\mu^2}{Q^2} \right) 
{\widetilde {\cal F}}_{n-k} \ ,
\label{tFnnu}
\ee
where we denote throughout ${\widetilde {\cal F}}_k \equiv {\widetilde {\cal F}}_k(1)$.

Since the coefficients $\tk_1(\nu_0+n)$ are RScl independent, the relations
(\ref{tF1})-(\ref{tF4}) are valid at any RScl $\mu^2$, i.e.,
we can replace there ${\widetilde {\cal F}}_j \mapsto {\widetilde {\cal F}}_j(\mu^2/Q^2)$
and ${\cal F}_j \mapsto {\cal F}_j(\mu^2/Q^2)$.

\subsection{Derivation of expressions for $k_m(\nu)$}
\label{App1:kmsder}

In this Subsection, we will omit the subscript pt ($a_{\rm pt} \mapsto a$).
Let's consider the relation
\be
\ta_{n+1}(Q^2) = \left(\frac{(-1)}{n\beta_0} \,   
\frac{d}{d\left( \ln Q^2 \right)}\right) \, \ta_{n}(Q^2) ,
\label{tarec}
\ee
which is a consequence of the definition (\ref{tan}).
In addition, we will use the following property:
\be
 \left(\frac{(-1)}{\beta_0} \,   
\frac{d}{d\left( \ln Q^2 \right)}\right) \, a^m ~=~
m \; a^{m+1} \, \left(1+ c_1 a + c_2 a^2 +c_3 a^3 +c_4 a^4 + \dots \right) \ ,
\label{r1.6}
\ee
which follows from the RGE, equation (\ref{RGE}).
We then obtain
\bea
\left(\frac{(-1)}{n\beta_0} \,   
\frac{d}{d\left( \ln Q^2 \right)}\right) \, \ta_{n}(Q^2) 
&=& a^{n+1} \, \left(1 + {\cal K}_1(n) a^{1} + {\cal K}_2(n) a^{2} + 
{\cal K}_3(n) a^{3} +{\cal K}_4(n) a^{4} + \dots \right) \nonumber \\ 
&\cdot & \left(1+ c_1 a + c_2 a^2 + c_3 a^3 +c_4 a^4 + \dots \right) \, ,
\label{r1.7}
\eea
where
\be
{\cal K}_j(n) ~=~ \frac{n+j}{n} \, k_j(n)
\label{r1.8}
\ee
Equation (\ref{r1.6}) can be rewritten as
\bea
\ta_{n+1}(Q^2)
&=& a^{n+1} \, \biggl(1 + \left[{\cal K}_1(n) + c_1\right] a^{1} + 
\left[{\cal K}_2(n) + c_1{\cal K}_1(n) + c_2\right] a^{2} 
\nonumber \\
&+& 
\left[{\cal K}_3(n) +  c_1{\cal K}_2(n) + c_2{\cal K}_1(n) + c_3\right]
a^{3} \nonumber \\
&+& 
\left[{\cal K}_4(n) +  c_1{\cal K}_3(n) +  c_2{\cal K}_2(n) + 
c_3{\cal K}_1(n) + c_4\right]
a^{3}
 + \dots \biggr) \, .
\label{r1.89}
\eea
Comparing equations (\ref{tanuanu}) and (\ref{r1.89}) we obtain the following 
recursive relations:
\bea
k_1(n+1) &=& \frac{(n+1)}{n} k_1(n) + c_1 \, ,  \label{r1.9} \\
k_2(n+1) &=&  \frac{(n+2)}{n} k_2(n) + c_1 \frac{(n+1)}{n} k_1(n) + 
c_2 \, , \label{r1.10} \\
k_3(n+1) &=& \frac{(n+3)}{n} k_3(n) +  c_1 \frac{(n+2)}{n} k_2(n) + 
c_2 \frac{(n+1)}{n} k_1(n) + c_3 , 
\label{r1.11} \\ 
k_4(n+1) &=& \frac{(n+4)}{n} k_4(n) +  c_1 \frac{(n+3)}{n} k_3(n) 
+  c_2 \frac{(n+2)}{n}k_2(n) + c_3 \frac{(n+1)}{n} k_1(n) + c_4 \, .
\label{r1.12}
\eea

We note that
\be
k_j(2)=c_j \ , \quad k_j(1) ~=~ 0 \, .
\label{r1.12a}
\ee

In order to solve the recursion relations (\ref{r1.9})-(\ref{r1.12}),
we find first solution to the following general recursion relation:
\be
k(n+1) ~=~ \frac{n+\alpha_1}{n+\alpha_2} \, k(n) + c(n) \ ,
\label{r2.1}
\ee
where $c(n)$ is some function of $n$ and $\alpha_1$ and $\alpha_2$ are some 
parameters.

If $c(n)=0$, then the solution of equation (\ref{r2.1}) is very simple
\be
k(n) ~=~ \widetilde{c}  \, \frac{\Gamma(n+\alpha_1)}{\Gamma(n+\alpha_2)} \ , 
\label{r2.2}
\ee
where $\widetilde{c}$ is an arbitrary constant and $\Gamma$ is the
Riemannian $\Gamma$-function.

If $c(n)\neq 0$, it is convenient to introduce a new variable $\hat{k}(n)$
which is related with $k(n)$ as
\be
k(n) ~=~ \frac{\Gamma(n+\alpha_1)}{\Gamma(n+\alpha_2)} \, \hat{k}(n) \ . 
\label{r2.3}
\ee
Using equation (\ref{r2.3}) in equation (\ref{r2.1}) we obtain
\be
\hat{k}(n+1) ~=~ \hat{k}(n) + 
\frac{\Gamma(n+\alpha_2+1)}{\Gamma(n+\alpha_1 +1)} \, c(n) \ ,
\label{r2.4}
\ee
with the solution
\be
\hat{k}(n) ~=~ \hat{k}(s) + \sum_{j=s}^{n-1}
\, \frac{\Gamma(j+\alpha_2+1)}{\Gamma(j+\alpha_1 +1)} \, c(j) \ ,
\label{r2.5}
\ee
where $s$ is a chosen number.
Below it will be convenient to use $s=2$, because $k_j(2)=c_j$.

So, for $r(n)$, we have
\bea
k(n) &=&  \frac{\Gamma(n+\alpha_1)}{\Gamma(n+\alpha_2)}
\left[\frac{\Gamma(s+\alpha_2)}{\Gamma(s+\alpha_1)} \, k(s) + 
\sum_{j=s+1}^{n}
\, \frac{\Gamma(j+\alpha_2)}{\Gamma(j+\alpha_1) } c(j-1)\right] ,
\nonumber\\
&=&  \frac{\Gamma(n+\alpha_1)}{\Gamma(n+\alpha_2)}
\left[\frac{\Gamma(2+\alpha_2)}{\Gamma(2+\alpha_1)} \, k(2) + 
\sum_{j=3}^{n}
\, \frac{\Gamma(j+\alpha_2)}{\Gamma(j+\alpha_1) } c(j-1)\right] ,
\label{r2.6}
\eea

Having the solution (\ref{r2.6}) to the recursion relation (\ref{r2.1}), we
can proceed to solving the recursion relations (\ref{r1.9})-(\ref{r1.12}).

{\bf 1.}~~ Consider the recursion (\ref{r1.9}): 
it corresponds to the general case with 
\be
k=k_1, \quad \alpha_1=1,~~  \alpha_2=0,~~ c(n)=c_1,~~ k_1(2)=c_1 \ .
\label{r2.7}
\ee
So, using the solution (\ref{r2.6}), and $k_1(2)=c_1$, we have
\be
k_1(n) ~=~ n c_1 \left( \frac{1}{2} + \sum^n_{j=3} \frac{1}{j} \right)
~=~  n c_1 \left(S_1(n)-1\right) \,
\label{r2.8a}
\ee
where
\be
S_m(n) ~=~ \sum^n_{j=1} \frac{1}{j^m}
\label{r2.9a}
\ee
are harmonic numbers (of $m$'th order), 
which can be related with the $(m-1)$'th derivative of 
$\Psi$-function:
\bea
S_{1}(n) &=& \Psi(n+1)-\Psi(1), 
\nonumber\\
S_{m+1}(n) &=& \frac{(-1)^m}{m!} \, 
\left( \Psi^{(m)}(n+1) - \Psi^{(m)}(1) \right), \quad
\Psi^{(m)}(x+1) ~=~ \frac{d^m}{dx^m} \Psi(x+1) \ .
\label{r2.10a}
\eea
The $\Psi$-function is in turn the logarithmic derivative of
the corresponding $\Gamma$-function:
\be 
\Psi(x+1) ~=~ \frac{d}{dx} \ln \Gamma(x+1)
\label{r2.10b}
\ee
and $\gamma=-\Psi(1)$ is Euler constant.

So, equation (\ref{r2.8a}) can be represented in the following form:
\be
k_1(n) ~=~  n c_1 \biggl(\Psi(n+1)-\Psi(1)-1\biggr) 
~=~  n c_1 \biggl(\Psi(n+1)-\Psi(2)\biggr) \, ,
\label{r2.8b}
\ee
which is well-defined also for noninteger values $n \mapsto \nu$
\footnote{It can be considered as the analytic continuation of the
coefficients $k_m(n)  \mapsto k_m(\nu)$ based on the corresponding procedure 
for harmonic numbers \cite{Kotikov:2005gr}.}
\be
k_1(\nu) ~=~  \nu c_1 \biggl(\Psi(\nu+1)-\Psi(2)\biggr) \, ,
\label{r2.8c}
\ee

{\bf 2.}~~ Consider the recursion (\ref{r1.10}): 
it corresponds to the general case
with 
\bea
\alpha_1 &=& 2,~~  \alpha_2=0,~~ k_2(2)=c_2,~~ \nonumber \\
c(n) &=& c_1 \cdot \frac{n+1}{n} k_1(n)
+c_2 = c_1^2 \cdot (n+1)[S_1(n)-1]+c_2
\, .
\label{r2.9b}
\eea

So, using the solution (\ref{r2.6}), we have
\be
k_2(n) ~=~ n(n+1) \left( \frac{1}{6} c_2 + \sum^n_{j=3} \frac{1}{j(j+1)} 
\biggl\{c_2+c_1^2 \, j[S_1(j-1)-1]\biggl\}\right)
\label{r2.10c}
\ee

The coefficient in front of $c_2$ has the form
\be
\frac{1}{6} + \sum^n_{j=3} \left(\frac{1}{j}-\frac{1}{j+1}\right)
~=~  \frac{1}{6} + \left(\sum^n_{j=3} - \sum^{n+1}_{j=4}\right) \frac{1}{j}
~=~ \frac{1}{6} + \left(\frac{1}{3} -\frac{1}{n+1} \right) ~=~
\frac{n-1}{2(n+1)} \ .
\label{r2.11}
\ee
The coefficient in front of $c_1^2$ has the form
\be
\sum^n_{j=3} \frac{1}{j+1} \, \biggl(S_1(j-1)-1\biggr)
~=~ \sum^n_{j=3} \left( \frac{S_1(j)}{j+1}-\frac{1}{j}\right)
\label{r2.12}
\ee
Here, the second term on the right-hand side is
\be
\sum^n_{j=3} \frac{1}{j} ~=~ S_1(n)-S_1(2) ~=~  S_1(n)- \frac{3}{2} \ ,
\label{r2.13}
\ee
while the first term is
\be
\sum^n_{j=3} \frac{S_1(j)}{j+1} ~=~ \sum^{n+1}_{j=4} \frac{S_1(j-1)}{j}
~=~ \frac{1}{2} \biggl( S_1^2(n+1)-S_2(n+1)\biggr) -1 ~\equiv
\frac{1}{2} Z_2(n+1) -1,
\label{r2.14}
\ee
where
\be
Z_m(n) ~=~ \frac{1}{\Gamma( n+1)} \frac{d^m}{(dx^m} 
{\left(\frac{\Gamma(n+1+x)}{ 
\Gamma(1+x)}\right)}{\Big |}_{x=0}
= \frac{d^m}{dx^m} \exp\left[\sum_{k=1}^{\infty} 
\frac{(-1)^{k+1} x^k}{k} S_k(n)\right]{\Big |}_{x=0}\, .
\label{r2.15}
\ee
For several first values, $Z_m(n)$ are
\bea
Z_0(n) &=& 1,~~~Z_1(n) ~=~ S_1(n),~~~
Z_2(n) ~=~ S_1(n)^2 - S_2(n) \, , 
\nonumber \\
Z_3(n) &=& S_1(n)^3 - 3 S_1(n) S_2(n) +2 S_3(n) \, , 
\nonumber \\
Z_4(n) &=& S_1(n)^4 - 6 S_1(n)^2 S_2(n) + 8S_1(n) S_3(n) + 
3 S_2(n)^2 - 6 S_4(n) \, .
\label{r2.15a}
\eea
In the case noninteger values $n \mapsto \nu$, the ($m$'th) order harmonic 
numbers (\ref{r2.9a}) are
\be
S_1(\nu) ~=~ \Psi(\nu+1)-\Psi(1), ~~
S_m(\nu) ~=~
\frac{(-1)^{m-1}}{(m-1)!} \, \left(\Psi^{(m-1)}(\nu+1)-\Psi^{(m-1)}(1)
\right) \, ,
\label{r2.15b}
\ee
where
\be
\Psi(\nu) ~=~ \frac{d}{(d \nu)} \ln \left(\Gamma(\nu)\right),~~~
\Psi^{(m-1)}(\nu) ~=~ \frac{d^m}{d \nu^m} \Psi(\nu)
\label{r2.15c}
\ee

So, the result for $k_2(n)$ has the form
\be
k_2(n) ~=~ \frac{n(n-1)}{2} \, c_2 + \frac{n(n+1)}{2} \, c_1^2
\cdot \left( Z_2(n+1) -2Z_1(n)+1 \right) \ ,
\label{r2.16a}
\ee
which is well-defined also for noninteger values $n \mapsto \nu$
\be
k_2(\nu) ~=~ \frac{\nu (\nu-1)}{2} \, c_2 + \frac{\nu(\nu+1)}{2} \, c_1^2
\cdot \left( Z_2(\nu+1) -2Z_1(\nu)+1 \right) \ .
\label{r2.16b}
\ee

{\bf 3.}~~ The results for $k_3(n)$ and $k_4(n)$ can be obtained similarly. 
After the replacement $n \mapsto \nu$ they have the forms as given
in equations (\ref{kmsk3})-(\ref{kmsk4}) and (\ref{Baux}).

\section{Large-$n$ behavior of coefficients}
\label{App2}

Here we calculate the asymptotical results for the coefficients
${\widetilde {\cal F}}_n$ at $n \to \infty$  of the expansion (\ref{Fmpta}),
assuming the standard 
Lipatov-type behavior \cite{Lipatov:1976ny} ${\cal F}_n \sim n!$ for the 
coefficients ${\cal F}_n$ of the original expansion (\ref{calFpt}).

It is convenient to use the following form for the coefficients ${\cal F}_n$ 
at $n \to \infty$:
\be
{\cal F}_n = \frac{\Gamma(n+\nu_0)}{\Gamma(\nu_0)} b^n \,,
\label{F1}
\ee
where $b \sim 1$ and $\Gamma$ is the Euler Gamma function.

From equations (\ref{tF1})-(\ref{tF4}) we conclude that
\be
{\widetilde {\cal F}}_n = \sum_{m=0}^{n} \tk_{n-m}(\nu_0+m){\cal F}_m,~~
\tk_{0}=1,~~ {\cal F}_0=1 \, .
\label{F2}
\ee

Firstly consider the first several terms on the right-hand side.

Let $m=n-1$. Due to equations (\ref{tkmsk1}) and (\ref{tBaux}), 
the  right-hand side of (\ref{F2}) has the following form:
\be
\tk_{1}(\nu_0+n-1){\cal F}_{n-1} = -c_1 \nu \Bigl[\tZ_1(\nu) - 1 \Bigr] \,
 \frac{\Gamma(\nu)}{\Gamma(\nu_0)} b^{n-1} {\bigg |}_{\nu=\nu_0+n-1}=  
-c_1  b^{n-1} 
\frac{\Gamma(n+\nu_0)}{\Gamma(\nu_0)} \Bigl[\tZ_1(\nu_0+n-1) - 1 \Bigr] \,.
\nonumber 
\ee
It can be rewritten also as
\be
\tk_{1}(\nu_0+n-1)\frac{{\cal F}_{n-1}}{{\cal F}_{n}} =
- \frac{c_1}{b}  \Bigl[\tZ_1(\nu_0+n-1) - 1 \Bigr] = - \frac{c_1}{b}
\left(\frac{d}{d x} -1 \right) \, 
\left( \frac{\Gamma(1-x)\Gamma(\nu_0+n)}{\Gamma(\nu_0+n-x)}
\right){\Big |}_{x=0} \ ,
\nonumber 
\ee
where we use equation (\ref{tZm}) on the right-hand side.

At $m=n-2$ we have from (\ref{tkmsk2}) and (\ref{tBaux})
\be
\tk_{2}(\nu_0+n-2)\frac{{\cal F}_{n-2}}{{\cal F}_{n}} = \frac{1}{b^2}
\bigg(- c_2 \, \tB_2(\nu_0+n-2) +  \frac{c_1^2}{2} \, \tB_{1,1}(\nu_0+n-2) 
\biggr) \, ,
\nonumber 
\ee
where at $n \to \infty$ (see equation (\ref{tBaux}))
\be
\tB_2(\nu_0+n-2) \approx \frac{1}{2} \ , \quad
\tB_{1,1}(\nu_0+n-2) \approx  \tZ_2(\nu_0+n-1) - 2  \, \tZ_1(\nu_0+n-1) + 1 \ .
\nonumber 
\ee
We use the simbol $\approx$ to show the asymptotics at $n \to \infty$.
In particular, in the $\tB_{1,1}$-case, the symbol  $\approx$ involves
the replacement of the argument $\nu_0+n-2$ in $\tZ_2$ by $\nu_0+n-1$.\footnote{
It can be checked that $\tZ_k(n+m)/\tZ_k(n) \to 1$ when $n \to \infty$.}
Similar replacements will be used below.

So, we have
\be
\tk_{2}(\nu_0+n-2)\frac{{\cal F}_{n-2}}{{\cal F}_{n}}  \approx
\frac{1}{2b^2} \left[ c_1^2 \left(\frac{d}{d x} -1 \right)^2 -c_2 \right]
\left( \frac{\Gamma(1-x)\Gamma(\nu_0+n)}{\Gamma(\nu_0+n-x)}
\right){\Big |}_{x=0} \ .
\nonumber 
\ee

When $m=n-3$, using (\ref{tkmsk3}) and (\ref{tBaux}) we have
\be
\tk_{3}(\nu_0+n-3)\frac{{\cal F}_{n-3}}{{\cal F}_{n}} = \frac{1}{2b^3}
\bigg(- c_3 \, \tB_3(\nu_0+n-3) + c_1c_2 \, \tB_{1,2}(\nu_0+n-3) - 
 \frac{c_1^3}{3} \, \tB_{1,1,1}(\nu_0+n-3) \biggr) \ ,
\nonumber 
\ee
where at $n \to \infty$ (see equation (\ref{tBaux}))
\bea
\tB_3(\nu_0 + n -3) &\approx & \frac{1}{6}\ ,
\quad
\tB_{1,2}(\nu_0+n-3) \approx 
\left(\tZ_1(\nu_0+n-1) -\frac{1}{6}\right) \ , 
\nonumber \\
\tB_{1,1,1}(\nu_0+n-3) & \approx & \tZ_3(\nu_0+n-1)-3 \, \tZ_2(\nu_0+n-1) + 
3  \, \tZ_1(\nu_0+n-1) - 1 \ .
\nonumber 
\eea
Hence we have
\be
\tk_{3}(\nu_0+n-3)\frac{{\cal F}_{n-3}}{{\cal F}_{n}}  \approx \frac{1}{6b^3}
\bigg(- \frac{c_3}{2}+ 3 c_1c_2 \,  \left(\frac{d}{d x} -\frac{1}{6} \right)
- c_1^3 \,  \left(\frac{d}{d x} -1 \right)^3 \biggr)
\left( \frac{\Gamma(1-x)\Gamma(\nu_0+n)}{\Gamma(\nu_0+n-x)}
\right){\Big |}_{x=0} \ .
\nonumber 
\ee

Finally, when $m=n-4$, equations (\ref{tkmsk3}) and (\ref{tBaux}) lead to
\bea
\tk_{4}(\nu_0+n-4)\frac{{\cal F}_{n-4}}{{\cal F}_{n}} &=& \frac{1}{6b^4}
\bigg(- c_4 \, \tB_4(\nu_0+n-4) + c_2^2 \, \tB_{2,2}(\nu_0+n-4) + 
\frac{c_1 c_3}{2} \, \tB_{1,3}(\nu_0+n-4)\nonumber \\
&&  -
\frac{c_1^2 c_2}{2} \,  \tB_{1,1,2}(\nu_0+n-4)  +\frac{c_1^4}{4} \, 
\tB_{1,1,1,1}(\nu_0+n-4) \biggr),
\eea
where at $n \to \infty$ (see equation (\ref{tBaux}))
\bea
\tB_4(\nu_0+n-4) &\approx  & \frac{1}{12}\ ,\quad
\tB_{2,2}(\nu_0+n-4)  \approx  \frac{13}{12}  \ ,\quad
\tB_{1,3}(\nu_0+n-4)  \approx  \tZ_1(\nu_0+n-1)+ \frac{1}{6} \ ,
\nonumber \\
\tB_{1,1,2}(\nu_0+n-4) &\approx & 3 \tZ_2(\nu_0+n-1) - \tZ_1(\nu_0+n-1)
+ \frac{8}{3} \ ,
\nonumber \\
\tB_{1,1,1,1}(\nu_0+n-4) &\approx &  
\tZ_4(\nu_0+n-1) - 4 \, \tZ_3(\nu_0+n-1) + 6 \,
\tZ_2(\nu_0+n-1) - 4  \, \tZ_1(\nu_0+n-1) + 1 \ .
\nonumber
\eea

So, we have
\bea
\tk_{4}(\nu_0+n-4)\frac{{\cal F}_{n-4}}{{\cal F}_{n}} &\approx & \frac{1}{24b^4}
\bigg( \frac{13 c_2^2-c_4}{3} + 
2c_1 c_3 \,  \left(\frac{d}{d x} +\frac{1}{6} \right) -
6 c_1^2 c_2 \, \left[\left(\frac{d}{d x} -\frac{1}{6} \right)^2 + \frac{31}{36}
\right] \nonumber \\
&&
+ c_1^4 \,  \left(\frac{d}{d x} -1 \right)^4 \biggr)
\left( \frac{\Gamma(1-x)\Gamma(\nu_0+n)}{\Gamma(\nu_0+n-x)}
\right){\Big |}_{x=0} \ .
\label{F2a}
\eea

\subsection{Contributions of the powers of $c_1$}

Taking only the terms $\sim c_1^m$ $(m=1,...,n)$, we have
\be
{\widetilde {\cal F}}_n^{(1)} = {\cal F}_n \sum^n_{m=0} 
\frac{(-1)^m c_1^m}{m! \; b^m}
\,  \left(\frac{d}{d x} -1 \right)^m 
\left( \frac{\Gamma(1-x)\Gamma(\nu_0+n)}{\Gamma(\nu_0+n-x)}
\right){\Big |}_{x=0} \ .
\label{F1.1}
\ee
At the beginning it is convenient to consider the sum on the 
right-hand side at 
$n \to \infty$. Moreover, the last term can be represented as
\be
\frac{\Gamma(1-x)\Gamma(\nu_0+n)}{\Gamma(\nu_0+n-x)} = (\nu_0+n-1) \,
\int_0^1 dy y^{-x} (1-y)^{\nu_0+n-2} \, .
\nonumber 
\ee
Application of the operator $(d/dx -\beta)$ to the term $y^{-x}$ has the simple
form
\be
\left(\frac{d}{d x} -\beta \right)^m \,  y^{-x} = \left(\ln \frac{1}{y} - 
\beta \right)^m \, .
\nonumber 
\ee
Then we have for the series on the right-hand side 
of (\ref{F1.1}) at $n \to \infty$
\be
\sum^n_{m=0} \frac{(-1)^m c_1^m}{m! b^m} \,   \left(\ln \frac{1}{y} - 
\beta \right)^m  = \exp \left[-\frac{c_1}{b} \left(\ln \frac{1}{y} - 
\beta \right)\right] = y^{c_1/b} e^{c_1\beta/b} \, .
\nonumber 
\ee
Thus, the contribution ${\widetilde {\cal F}}_n^{(1)}$ has the following form
after integration on $y$:
\be
{\widetilde {\cal F}}_n^{(1)} = {\cal F}_n R_n \left[e^{c_1/b}
 \frac{\Gamma(1+c_1/b)\Gamma(\nu_0+n)}{\Gamma(\nu_0+n+c_1/b)}\right] \, ,
\nonumber 
\ee
where the operation $R_n[F(c_1)]$ takes the first $(n+1)$ terms of the expansion 
of $F$ in powers of $c_1$, and $R_n(x) \to x$ when $n \to \infty$.

Since $n$ is large, the difference between $R_n[F(c_1)]$  
and $F$ is very small ($\sim 1/(n+1)!$). 
So, we can omit the $R_n$ operation and take the contribution of
 the terms $\sim c_1^m$ $(m=1,...,n)$ in the form
\be
{\widetilde {\cal F}}_n^{(1)} \approx {\cal F}_n e^{c_1/b}
 \frac{\Gamma(1+c_1/b)\Gamma(\nu_0+n)}{\Gamma(\nu_0+n+c_1/b)} \, .
\label{F1.3}
\ee

Note that the contribution ${\widetilde {\cal F}}_n^{(1)}$ is very important 
because the coefficient $c_1$ is universal and has nonzero value
in ${\rm MS}$-like schemes. In the scheme where all $c_j=0$ $(j\geq 2)$,
${\widetilde {\cal F}}_n^{(1)}$ represents the full contribution to the 
reexpression of the expansion (\ref{calFpt}) to the one 
of equation (\ref{Fmpta}): ${\widetilde {\cal F}}_n^{(1)} = {\widetilde {\cal F}}_n$. 
As we can see
in the next subsection, the contributions $\sim c_j$ $(j\geq 2)$ can be 
expressed also in the form $\sim {\widetilde {\cal F}}_n^{(1)}$.

Use of Stirling's formula in (\ref{F1.3}) gives us
\be
{\widetilde {\cal F}}_n^{(1)}/{\cal F}_n \approx
e^{c_1/b}  \Gamma(1 + c_1/b) \frac{1}{n^{c_1/b}} \ .
\label{Sitrl}
\ee
When the coefficients in the original series are nonalternating in
sign, such as the one encountered
in the Higgs decay width, equation (\ref{calFn}), $b$ is positive
and the above ratio even tends to zero when $n$ increases (we note that
$c_1$ is positive for all $n_f \leq 6$).

Thus, the Lipatov-type asymptotics takes place for both ${\cal F}_n$ and
${\widetilde {\cal F}}_n $ coefficients in the $c_j=0$ ($j \geq 2$) scheme: only 
the subasymptotical terms are changed.

\subsection{Contributions of $\sim c_j$  $(j\geq 2)$}

The terms proportional to the first power of $c_2$ have the following form:
\be
{\widetilde {\cal F}}_n^{(2)} \approx {\cal F}_n 
\bigg( -\frac{c_2}{2b^2} + \frac{c_1c_2}{2b^3}
\,  \left(\frac{d}{d x} -\frac{1}{6} \right) - \frac{c_1^2c_2}{4b^4}
\, \left[\left(\frac{d}{d x} -\frac{1}{6} \right)^2 + \frac{31}{36}
\right] + ...
 \biggr)
\left( \frac{\Gamma(1-x)\Gamma(\nu_0+n)}{\Gamma(\nu_0+n-x)}
\right){\Big |}_{x=0} \ .
\nonumber 
\ee
It is convenient at the first stage to exclude the term $\sim (31/36) c_1^2c_2$
from the consideration. It will be considered below together with the term
$\sim c_4$.

Without the term $\sim (31/36) c_1^2c_2$, the contribution 
${\widetilde {\cal F}}_n^{(2)}$ has the following form:
\be
{\widetilde {\cal F}}_n^{(2)} = -\frac{c_2}{2b^2} \, {\cal F}_n 
\sum^{n-2}_{m=0} 
\frac{(-1)^m c_1^m}{m! \; b^m}
\,  \left(\frac{d}{d x} -\frac{1}{6} \right)^m 
\left( \frac{\Gamma(1-x)\Gamma(\nu_0+n)}{\Gamma(\nu_0+n-x)}
\right){\Big |}_{x=0} \ .
\label{F2.1}
\ee

Repeating the calculations for ${\widetilde {\cal F}}_n^{(1)}$ done in the 
previous subsection we obtain for ${\widetilde {\cal F}}_n^{(2)}$
\be
{\widetilde {\cal F}}_n^{(2)} \approx -\frac{c_2}{2b^2} \,{\cal F}_n R_{n-2} 
\left[e^{c_1/(6b)}
 \frac{\Gamma(1+c_1/b)\Gamma(\nu_0+n)}{\Gamma(\nu_0+n+c_1/b)}\right] \, ,
\nonumber 
\ee
where, as in the previous subsection, 
the operation $R_{n-2}[F(c_1)]$ takes the first $(n-1)$ terms of the expansion 
of $F$ on $c_1$.

Since $n$ is large, the difference between $R_{n-2}[F(c_1)]$  and $F$ is 
very small. So, we can omit the $R_{n-2}$ operation and take the 
contribution of the terms $\sim c_1^m$ $(m=1,...,n-2)$ in the form
\be
{\widetilde {\cal F}}_n^{(2)} \approx -\frac{c_2}{2b^2} {\cal F}_n e^{c_1/(6b)}
 \frac{\Gamma(1+c_1/b)\Gamma(\nu_0+n)}{\Gamma(\nu_0+n+c_1/b)} 
\approx  -\frac{c_2}{2b^2} {\widetilde {\cal F}}_n^{(1)}   e^{-5c_1/(6b)}
\, .
\label{F2.2}
\ee

Taking the terms  $\sim c_3$, we have
\bea
{\widetilde {\cal F}}_n^{(3)} &\approx & {\cal F}_n 
\bigg( -\frac{c_3}{12b^3} + \frac{c_1c_3}{12b^4}
\,  \left(\frac{d}{d x} +\frac{1}{6} \right)  + ...
 \biggr)
\left( \frac{\Gamma(1-x)\Gamma(\nu_0+n)}{\Gamma(\nu_0+n-x)}
\right){\Big |}_{x=0} \nonumber \\
&=& -\frac{c_3}{12b^3} \, {\cal F}_n 
\bigg(1 +  \frac{(-c_1)^1}{1! \; b^1}
\,  \left(\frac{d}{d x} +\frac{1}{6} \right)  + ...
 \biggr)
\left( \frac{\Gamma(1-x)\Gamma(\nu_0+n)}{\Gamma(\nu_0+n-x)}
\right){\Big |}_{x=0}\ .
\nonumber 
\eea

Repeating the above calcuations, we obtain
\be
{\widetilde {\cal F}}_n^{(3)} \approx -\frac{c_3}{12b^3} \,{\cal F}_n R_{n-3} 
\left[e^{-c_1/(6b)}
 \frac{\Gamma(1+c_1/b)\Gamma(\nu_0+n)}{\Gamma(\nu_0+n+c_1/b)}\right] \ , 
\nonumber 
\ee
and,  because $n$ is large,
\be
{\widetilde {\cal F}}_n^{(3)} \approx  -\frac{c_3}{12b^3} 
{\widetilde {\cal F}}_n^{(1)}   e^{-7c_1/(6b)}
\, .
\label{F2.3}
\ee

Now we consider the remaining terms $\sim 1/b^4$. At the leading order (in $c_1$),
they contribute in two places:
as the term $\sim (13 c_2^2-c_4)$ in equation (\ref{F2a}) and as 
the term $\sim (31/36) c_1^2c_2$ in the first equation of this subsection.
Taking them together we have, at the leading order in $c_1$
\be
{\widetilde {\cal F}}_n^{(4)} \approx  - 
\frac{(2c_4 -26 c_2^2 + 31  c_1^2c_2)}{144 b^4} \, {\cal F}_n \ .
\nonumber 
\ee
Adding to this the terms of relative higher order in $c_1$,
in analogy with above calculations for 
${\widetilde {\cal F}}_n^{(j)}$ $(j=2,3)$, gives
\be
{\widetilde {\cal F}}_n^{(4)} \approx  
- \frac{(2c_4 -26 c_2^2 + 31  c_1^2c_2)}{144 b^4} \,
{\widetilde {\cal F}}_n^{(1)}   e^{-(\hat{k} +1)c_1/b}
\, .
\label{F2.4}
\ee 
To find the exact value of the factor $\hat{k}$ we should calculate the term 
$\tk_{5}(\nu_0+n-5){\cal F}_{n-5}/{\cal F}_{n}$ in analogy with 
(\ref{F2a}). It needs in turn
the calculaton of the coefficients $\tk_{5}$ and, thus, one step more 
in the analysis in appendix \ref{App1}. 
Hovewer, looking carefully at the above calculations, we note that
the coefficients in the exponents, in front of $-c_1/b$, 
rise with the index $j$ of ${\widetilde {\cal F}}_n^{(j)}$ . For 
${\widetilde {\cal F}}_n^{(3)}$ the corresponding coefficient is equal to $7/6$
and we suggest that $\hat{k}$ in (\ref{F2.4}) should be bigger.

So, we have for the coefficient ${\widetilde {\cal F}}_n$ the following
approximation at large $n$ values:
\be
{\widetilde {\cal F}}_n \approx  
{\widetilde {\cal F}}_n^{(1)}  \,
\left(1 -\frac{c_2}{2b^2} e^{-5c_1/(6b)} -\frac{c_3}{12b^3} e^{-7c_1/(6b)}
- \frac{(2c_4 -26 c_2^2 + 31  c_1^2c_2)}{144 b^4} \, e^{-(\hat{k} +1)c_1/b} 
- ... \right)
\, .
\label{F2.5}
\ee 
We see that the corrections from $\sim c_j$  $(j\geq 2)$ have the same sign
and are decreasing in magnitude. Indeed, for $b=1$, and
when $n_f=5$, we have
\be
b=1,~~ c_1=1.2609,~~c_2=1.4748,~~c_3=9.8357,~~c_4 \approx 86. \ ,
\nonumber
\ee
and these corrections apparently have decreasing magnitudes
\be
{\widetilde {\cal F}}_n \approx  
{\widetilde {\cal F}}_n^{(1)}  \,
\left( 1 - 0.258 - 0.188 - 1.3 \times (0.283)^{(\hat{k} +1)} 
- ... \right)
\, ,
\label{F2.6}
\ee 
where  $\exp[-(\hat{k} +1)c_1] \approx (0.283)^{(\hat{k} +1)}  < 0.23$ if $\hat{k} \geq 1/6$.

Thus, the contributions of $\sim c_j$  $(j\geq 2)$ are rather small and can be
expressed through the contribution of $\sim c_1$.

\section{Proof of equation (\ref{tHnun1})}
\label{App3}

We prove the formula of equation (\ref{tHnun1}) by mathematical induction
with respect to $n=0,1,\ldots$. For $n=0$, this is the formula of 
equation (\ref{tHnu1}) which was proven in the text. Now suppose that the
formula equation (\ref{tHnun1}) is valid for a given $n$. We will show that
then it must be valid also for $n+1$.

If the formula of equation (\ref{tHnun1}) is valid for a given $n$, we can use it
and the recursion relation (\ref{tHrecur}) to obtain
\footnote{A somewhat similar procedure, in the context of one-loop fractional perturbation theory, was performed in \cite{Kotikov:2010bm}
(see appendix A there).}
\bea
\tH_{\delta + n + 2}(\sigma) & = & 
\frac{(-1) \sin( \pi (\delta + n))}{\pi^2 (\delta+n+1) (\delta+n) \beta_0^{\delta+n+1}}
\nonumber\\
&& \times
\int_{\epsilon}^{\infty} \frac{d w}{w^{\delta+n}} 
\left[
\frac{\partial \rho_1(\sigma e^w)}{\partial w} 
- \frac{d \rho_1(\sigma)}{d \ln \sigma} 
- \frac{w}{1!} \frac{d^2 \rho_1(\sigma)}{d (\ln \sigma)^2}
 - \ldots
-  \frac{w^{n-1}}{(n-1)!} \frac{d^n \rho_1(\sigma)}{d (\ln \sigma)^n} 
\right] 
\label{tHnun2a}
\\
& = &
\frac{\sin( \pi (\delta + n + 1))}{\pi^2 (\delta+n+1) \beta_0^{\delta+n+1}}
{\bigg\{} 
\int_{\epsilon}^{\infty} \frac{d w \rho_1(\sigma e^w)}{w^{\delta+n+1}}
- \frac{1}{(\delta+n) \epsilon^{\delta+n}} \rho_1(\sigma e^{\epsilon})
- \frac{1}{(\delta+n)(\delta+n-1) \epsilon^{\delta+n-1}}
\frac{d \rho_1(\sigma)}{d \ln \sigma} 
\nonumber\\
&& -  \frac{1}{(\delta+n)(\delta+n-2) 1! \; \epsilon^{\delta+n-2}}
\frac{d^2 \rho_1(\sigma)}{d (\ln \sigma)^2} - \cdots 
-  \frac{1}{(\delta+n)(\delta+n-k) (k-1)! \; \epsilon^{\delta+n-k}}
\frac{d^k \rho_1(\sigma)}{d (\ln \sigma)^k} - \cdots 
\nonumber\\
&&
- \frac{1}{(\delta+n)(\delta) (n-1)! \; \epsilon^{\delta}}
\frac{d^n \rho_1(\sigma)}{d (\ln \sigma)^n} {\big\}} \ .
\label{tHnun2b}
\eea
Here it is understood that $\epsilon \to +0$; and in the step from
equation (\ref{tHnun2a}) to equation (\ref{tHnun2b}) we performed integration
by parts in the first term.

Now we use in $\rho_1(\sigma e^{\epsilon})$ Taylor expansion in logarithm of
the argument ($\ln \sigma + \epsilon$)
\be
\rho_1(\sigma e^{\epsilon}) = \rho_1(\sigma) + 
\sum_{k=1}^{n} \frac{\epsilon^{k}}{k !}
\frac{d^{k} \rho_1(\sigma)}{(d \ln \sigma)^{k}} +
{\cal O}(\epsilon^{n+1}) \ ,
\label{Texp}
\ee
in the above expression (\ref{tHnun2b}), and obtain
\bea
\tH_{\delta + n + 2}(\sigma) & = & 
\frac{\sin( \pi (\delta + n + 1))}{\pi^2 (\delta+n+1) \beta_0^{\delta+n+1}}
{\bigg\{} \int_{\epsilon}^{\infty} \frac{d w \rho_1(\sigma e^w)}{w^{\delta+n+1}}
- \frac{\rho_1(\sigma)}{(\delta+n) \epsilon^{\delta+n}}
\nonumber\\
&&
- \sum_{k=1}^{n} \frac{1}{\epsilon^{\delta+n-k} (\delta+n)}
\left[ \frac{1}{k !} + \frac{1}{(\delta+n-k) (k-1)!} \right]
\frac{d^{k} \rho_1(\sigma)}{(d \ln \sigma)^{k}} + {\cal O}(\epsilon^{1-\delta})
{\bigg\}} \ .
\label{tHnun2c}
\eea
Using the identity
\be
\left[ \frac{1}{k !} + \frac{1}{(\delta+n-k) (k-1)!} \right] =
\frac{1}{k!} \frac{(\delta+n)}{(\delta+n-k)} \ ,
\label{idaux}
\ee
we can rewrite the expression (\ref{tHnun2c}) as
\bea
\tH_{\delta + n + 2}(\sigma) & = & 
\frac{\sin( \pi (\delta + n + 1))}{\pi^2 (\delta+n+1) \beta_0^{\delta+n+1}}
\int_{\epsilon}^{\infty} \frac{d w}{w^{\delta+n+1}}
{\bigg\{} \rho_1(\sigma e^w) - \rho_1(\sigma)
- \sum_{k=1}^{n} \frac{1}{k !} w^k  
\frac{d^{k} \rho_1(\sigma)}{(d \ln \sigma)^{k}} {\bigg\}}
+ {\cal O}(\epsilon^{1-\delta}) \ .
\label{tHnun2d}
\eea
Since $0 < \delta < 1$, we are now allowed to take the limit $\epsilon \to +0$
in the above integral ($\epsilon^{1-\delta} \to 0$) and we conclude that
the identity (\ref{tHnun1}) is valid also for $n+1$. This concludes the
proof of identity  (\ref{tHnun1}) via mathematical induction.

\section{Coefficients of expansion of ${\overline {\rm MS}}$ squared mass}
\label{App4}

Integration of the RGE's (\ref{RGE}) and (\ref{RGEm}) gives for the
${\overline {\rm MS}}$ squared running mass the solution in the form
of expansion (\ref{barm2run}), with the coefficients ${\cal M}_j$ there
being (note that $\gamma_0=1$)
\bea
{\cal M}_1 & = & - \frac{2}{\beta_0} ( c_1 - \gamma_1) \ ,
\label{cM1}
\\
{\cal M}_2 & = & \frac{1}{2}{\cal M}_1^2 
- \frac{1}{\beta_0} \left(  (c_2-\gamma_2) -c_1 (c_1-\gamma_1) \right) \ ,
\label{cM2}
\\
{\cal M}_3 & = & - \frac{1}{3} {\cal M}_1^3 + {\cal M}_1 {\cal M}_2
- \frac{2}{3 \beta_0} \left( (c_3-\gamma_3) -c_1 (c_2-\gamma_2) + 
(c_1^2-c_2)(c_1-\gamma_1) \right) \ ,
\label{cM3}
\\
{\cal M}_4 & = & \frac{1}{4} {\cal M}_1^4 - {\cal M}_1^2 {\cal M}_2
+ \frac{1}{2} {\cal M}_2^2 + {\cal M}_1 {\cal M}_3
\nonumber\\
&&
- \frac{1}{2 \beta_0} \left( (c_4-\gamma_4) -c_1 (c_3-\gamma_3) + 
(c_1^2-c_2)(c_2-\gamma_2) + (-c_1^3 + 2 c_1 c_2 - c_3)(c_1 - \gamma_1) \right) \ .
\label{cM4}
\eea

\end{document}